\documentclass[aps,prd,preprint,floatfix,superscriptaddress,nofootinbib,longbibliography,a4paper]{revtex4-1}
\pdfoutput=1
\usepackage{color}
\usepackage{graphicx}
\usepackage{textcomp}
\usepackage{subfigure}
\usepackage{feynmp}
\usepackage{amsmath,amssymb, array}
\usepackage{enumerate}
\usepackage{slashed}
\usepackage[normalem]{ulem}
\usepackage{hhline}
\usepackage{hyperref}
\usepackage{empheq}
\usepackage{cases}
\hypersetup{colorlinks=true,
    citecolor=blue,
	linkcolor=blue,
	filecolor=magenta,      
	urlcolor=blue}
\newcommand{\mathsym}[1]{{}}

\newcommand{\beqa}{\begin{eqnarray}}
\newcommand{\eeqa}{\end{eqnarray}}
\newcommand{\be}{\begin{equation}}
\newcommand{\ee}{\end{equation}}
\newcommand{\ba}{\begin{array}} 
\newcommand{\ea}{\end{array}}

\begin{document} 
\title{Hierarchies from deterministic non-locality in theory space Anderson localisation}
\bigskip
\author{Ketan M. Patel}
\email{ketan.hep@gmail.com}
\affiliation{Theoretical Physics Division, Physical Research Laboratory, Navarangpura, Ahmedabad-380009, India \vspace*{1cm}}

\begin{abstract}
The nearest-neighbour or local mass terms in theory space among quantum fields, with their generic disordered values, are known to lead to the localisation of mass eigenstates, analogous to Anderson localisation in a one-dimensional lattice. This mechanism can be used to create an exponential hierarchy in the coupling between two fields by placing them at opposite ends of the lattice chain. Extending this mechanism, we show that when copies of such fields are appropriately attached to the lattice chain, it leads to the emergence of multiple massless modes. These vanishing masses are a direct consequence of the locality of interactions in theory space. The latter may break down in an ordered and deterministic manner through quantum effects if additional interactions exist among the chain fields. Such non-locality can induce small masses for the otherwise massless modes without necessarily delocalising the mass eigenstates. We provide examples of interactions that preserve or even enhance localisation. Applications to flavour hierarchies, neutrino mass, and the $\mu$-problem in supersymmetric theories are discussed.
\end{abstract}

\maketitle
 
\section{Introduction}
\label{sec:intro}
The Standard Model (SM) is formulated as a quantum field theory (QFT) that involves a sizeable number of non-generic parameters. Some of these parameters are incalculable and exhibit hierarchical structures. Extensions of the SM, introduced to accommodate observationally established phenomena such as neutrino masses, dark matter etc., often exacerbate this issue by introducing additional hierarchical couplings and/or new scales into the theory. Understanding the origin of such hierarchies in couplings or scales \cite{Dirac:1937ti} has been a topic of considerable interest. An even more ambitious goal is to render at least some of these parameters calculable in terms of a small set of couplings, which can take only generic values. Several interesting mechanisms have been proposed to address these challenges, including dimensional transmutation \cite{Weinberg:1975gm,Susskind:1978ms} in four-dimensional QFT, wave-function localisation and/or warping in extra dimensions \cite{Grossman:1999ra,Gherghetta:2000qt,Abe:2002rj,Falkowski:2002cm,Randall:2002qr}, and four-dimensional theories featuring lattice-like structures in theory space \cite{Arkani-Hamed:2001nha,Arkani-Hamed:2001kyx,Hill:2000mu,Choi:2014rja,Giudice:2016yja,Kaplan:2015fuy,Craig:2017ppp}.

The mechanism of our interest in this work is the one proposed in \cite{Craig:2017ppp,Rothstein:2012hk}, which closely resembles the phenomenon of Anderson localisation \cite{Anderson:1958vr} in real lattice systems. The latter corresponds to the localisation of electron energy wavefunctions in a one-dimensional lattice due to strong disorder in the onsite energies in the lattice. Analogous localisation in the mass eigenstate of a quantum field arises when the interactions between several copies of such fields are arranged in a one-dimensional lattice-like form in the theory space, and strong disorder in the on-site couplings is introduced. The localisation can then be utilised to arrange small couplings or mass terms between fields by coupling them appropriately far enough on the lattice, as demonstrated in \cite{Craig:2017ppp}. There are primarily two features that make this mechanism very different from the other QFT space latticised models listed in \cite{Arkani-Hamed:2001nha,Arkani-Hamed:2001kyx,Hill:2000mu,Choi:2014rja,Giudice:2016yja,Kaplan:2015fuy}. Firstly, no elaborate ordering in the fundamental couplings are required. The localisation of the mass eigenstates emerges purely from the disorder. Therefore, the QFT couplings can be very generic. Secondly, there are no massless states. All the fields, including those at the edges of the lattice, have masses of the order of the size of the random couplings. This can also be attributed to the absence of any special arrangement other than the nearest-neighbour couplings or locality in the theory space, which is anyway a common feature characterizing the lattice-like structure of such frameworks.

In this work, we show that it is possible to obtain multiple massless modes within this class of models without compromising the characteristic features of the Anderson localisation mechanism. The massless states are not part of the one-dimensional lattice structure, but they are the ones that interact with the lattice fields in a specific manner. The vanishing of their mass is entirely due to the nearest neighbour, or local in theory space, interactions of the fields on the lattice. We show that the controlled non-local effects can generate masses for the otherwise massless modes. Such non-local effects can arise as quantum effects if the fields in the lattice chain have additional interactions, and they are completely deterministic or calculable in terms of the fundamental couplings. Interestingly, these effects do not destroy the Anderson localisation in general, and we point out some example scenarios in which the localisation remains intact or becomes even stronger after accounting for the radiative corrections. The induced masses for the otherwise massless modes are suppressed by both the loop factor and the exponential localisation. The phenomenological advantages of such arrangements are discussed in the context of some example applications.

The article is organised as follows. In the next section, we briefly review the original mechanism. Section \ref{sec:massless} discusses the arrangement that leads to massless modes. The non-local effects in the context of general abelian gauge interactions are computed in section \ref{sec:nonlocal}, followed by some explicit examples in section \ref{sec:examples}. We then discuss a few phenomenological applications of this construction in section \ref{sec:application} before concluding in section \ref{sec:concl}.

\section{Anderson localisation on fermion chain}
\label{sec:anderson}
\begin{figure}[t!]
\centering
\subfigure{\includegraphics[width=0.75\textwidth]{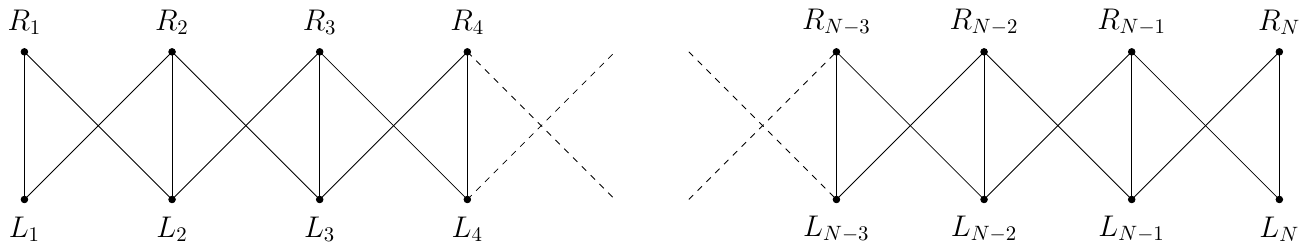}}
\caption{Fermionic chain in QFT space, giving rise to the Anderson localisation effect. The lattice sites denote left- or right-chiral fermions, while the edges denote the interactions between them.}
\label{fig1}
\end{figure}
The fermionic version of the QFT space Anderson localisation mechanism proposed in \cite{Craig:2017ppp} is as follows. Consider $N$ copies of Dirac fermions whose left- and right-chiral components are denoted by $L_i$ and $R_i$, respectively. The interactions between them are arranged in a one-dimensional lattice form in the theory space, such that the fermions at a given site have on-site mass terms and couplings with the nearest neighbours only. The chain is depicted in Fig. \ref{fig1} and the corresponding mass Lagrangian can be written as:
\beqa \label{LFC}
-{\cal L}_{\rm FC} &=& \sum_{i=1}^N\,\epsilon_i\,\overline{L}_i R_i+\sum_{i=1}^N\, t\,\left(\overline{L}_i R_{i+1}+\overline{L}_{i+1} R_{i}\right) +{\rm h.c.}\, \nonumber \\
& \equiv & \sum_{i,j}\,M_{ij}\,\overline{L}_i R_j+{\rm h.c.}\,,\eeqa
with 
\be \label{M_FC}
M_{ij} = \epsilon_{i}\,\delta_{ij} + t\, (\delta_{i+1,j} + \delta_{i,j+1})\,.\ee
The on-site mass terms, $\epsilon_i$,  are taken non-universal and they can fluctuate randomly. The off-site local couplings denoted by $t$ are assumed to be site-independent and isotropic. The mass matrix $M$ in eq. (\ref{M_FC}) is identical to the Hamiltonian of Anderson's tight-binding model \cite{Anderson:1958vr} with diagonal disorder.

In QFT, the Lagrangian in eq. (\ref{LFC}) can simply originate from a global symmetry containing $N$ factors of $U(1)$ group in a construction similar to ``quivers" or ``moose" in four dimensional theories \cite{Georgi:1985hf,Douglas:1996sw,Arkani-Hamed:2001kyx}. Under the $U(1)_j$, only the fermions at the $j^{\rm th}$ site are charged with vectorlike charges. The parameter $t$ can be seen as spurions which are charged under two nearest $U(1)$ factors. Its non-vanishing value breaks completely $U(1)^N$. In the limit of $t \to 0$ and $\epsilon_i \to 0$, the theory has a larger $U(1)^N_L \times U(1)^N_R$ global symmetry.

Assuming $\epsilon_i$ and $t$ as real parameters for simplicity, the matrix $M$ can be diagonalised using
\be \label{UMU}
U^T M U = {\rm Diag.}(m_1,...,m_N) \equiv D\,,\ee
with $m_i$ being real and positive. $U$ is a unitary matrix which can be written as $U= O P$ where $O$ is a real orthogonal matrix and $P$ is a diagonal matrix with elements $1$ or $i$. The $k^{\rm th}$ column of $U$, namely $v^{(k)}$, is an eigenvector of $M$ with eigenvalue $m_k$. In other words, 
\be \label{U_EV}
U = \begin{pmatrix}
v^{(1)} & v^{(2)}& ... & v^{(N)}
\end{pmatrix} \,.\ee
Starting from the original bases $L_i$ and $R_i$, the fermions in the physical basis are obtained by the rotations $(U^\dagger L)_i$ and $(U^\dagger R)_i$, respectively.

The most noteworthy aspect of the mass matrix $M$ in eq. (\ref{M_FC}) is that its eigenvectors are exponentially localised \cite{Anderson:1958vr}. This means that the elements of a given eigenvector can be approximated as
\be \label{vk_exp}
\left|v_j^{(k)}\right| \sim \left|v_{k_0}^{(k)}\right|\, e^{-\frac{|j-k_0|}{L_k}}\,,\ee
where $k_0$ is the site at which the eigenvector $v^{(k)}$ has its element with the largest magnitude.  As one moves away from $k_0$, the magnitude of the elements decreases exponentially, with the rate approximated in terms of the localisation length $L_k$. For sufficiently large $N$, the localisation length can be defined as an average over the logarithmic ratio of the nearby components of the eigenvector \cite{Izrailev_2012}, i.e.,
\be \label{L_k}
L_k^{-1} = -\frac{1}{N-1}\,\sum_{j=2}^N \ln\left| \frac{v_j^{(k)}}{v_{j-1}^{(k)}}\right|\,.\ee
The degree of localisation depends on the width of distribution, say $W$, from which the values of $\epsilon_i$ are randomly drawn and on the off-site coupling $t$. Localisation is observed in both $t < W$ and $t>W$ limits. The former leads to strong localisation, while the latter corresponds to a relatively weak localisation regime. Furthermore, the localisation persists even when one turns on the disorder in the nearest-neighbour couplings, as long as it is independent of the disorder in on-site terms \cite{PhysRevB.4.396}.

An analytical estimation of the localisation length in case of very strong disorder, i.e. $t/W \ll 1$, gives \cite{Izrailev_2012}
\be \label{Lk_th_1}
L_k^{-1} \simeq -\ln\left(\frac{2t}{W}\right)-1\,.\ee
Consequently, small $t/W$ leads to small localisation length and hence strong hierarchy among the elements of a given eigenvector through eq. (\ref{vk_exp}). A more accurate expression which fits the numerical simulations better can be derived from \cite{JLPichard_1986} and is given by
\be \label{Lk_th_2}
L_k^{-1} \simeq F(z_+)+F(z_-)\,,\ee
with $z_{\pm} = \frac{W}{2t}\pm\left(\frac{m_k}{t}-\frac{W}{2t}\right)$, and 
\be \label{Fz}
F(z) = \frac{2t}{W}\left(z\,\ln(z+\sqrt{z^2-4}) - \sqrt{z^2-4} - z\, \ln 2 \right)\,.\ee
Analytical estimations of localisation length in the weak disorder regime are also summarised in \cite{Izrailev_2012}.

\begin{figure}[t]
\centering
\subfigure{\includegraphics[width=0.42\textwidth]{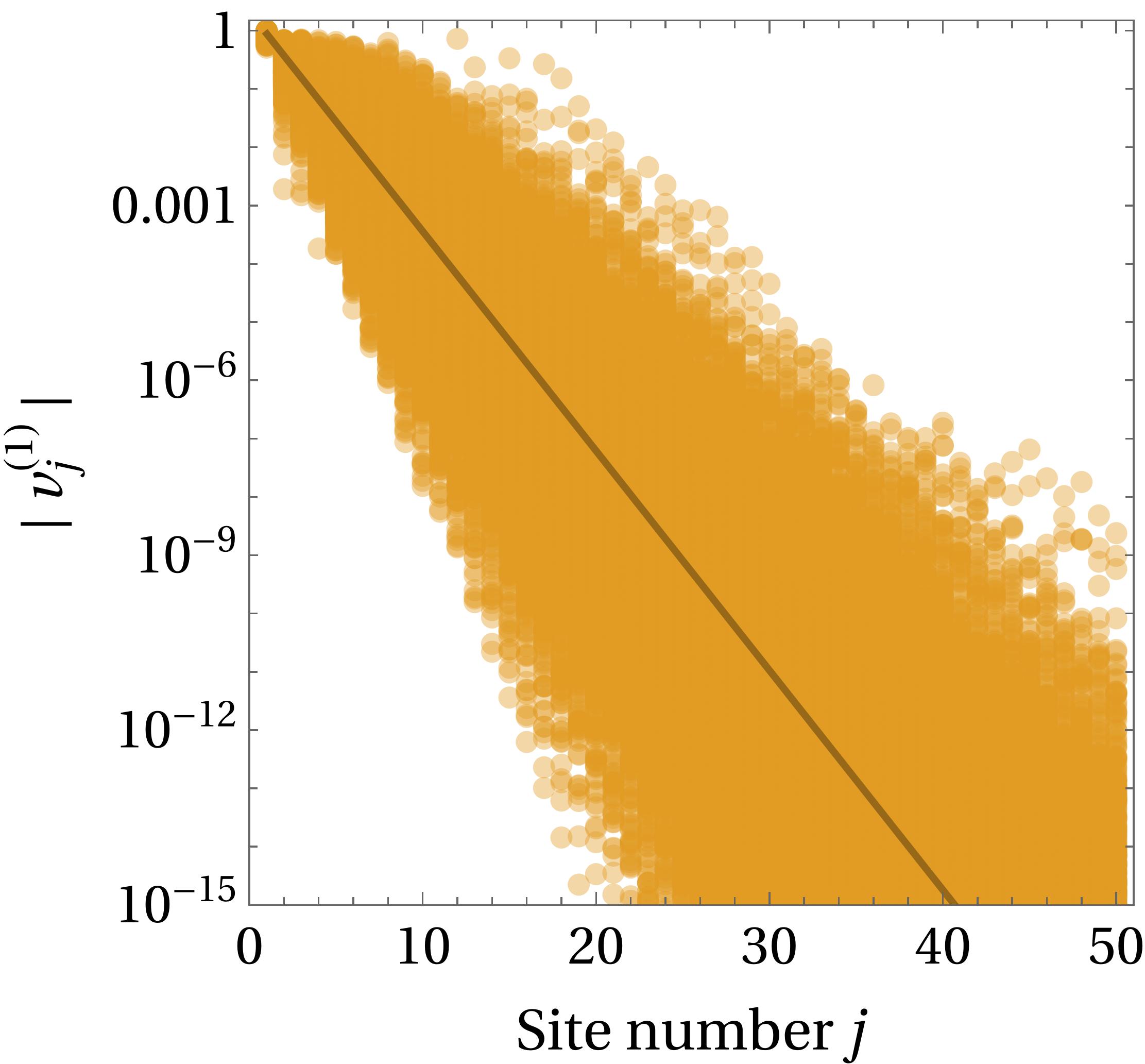}}\hspace*{1cm}
\subfigure{\includegraphics[width=0.42\textwidth]{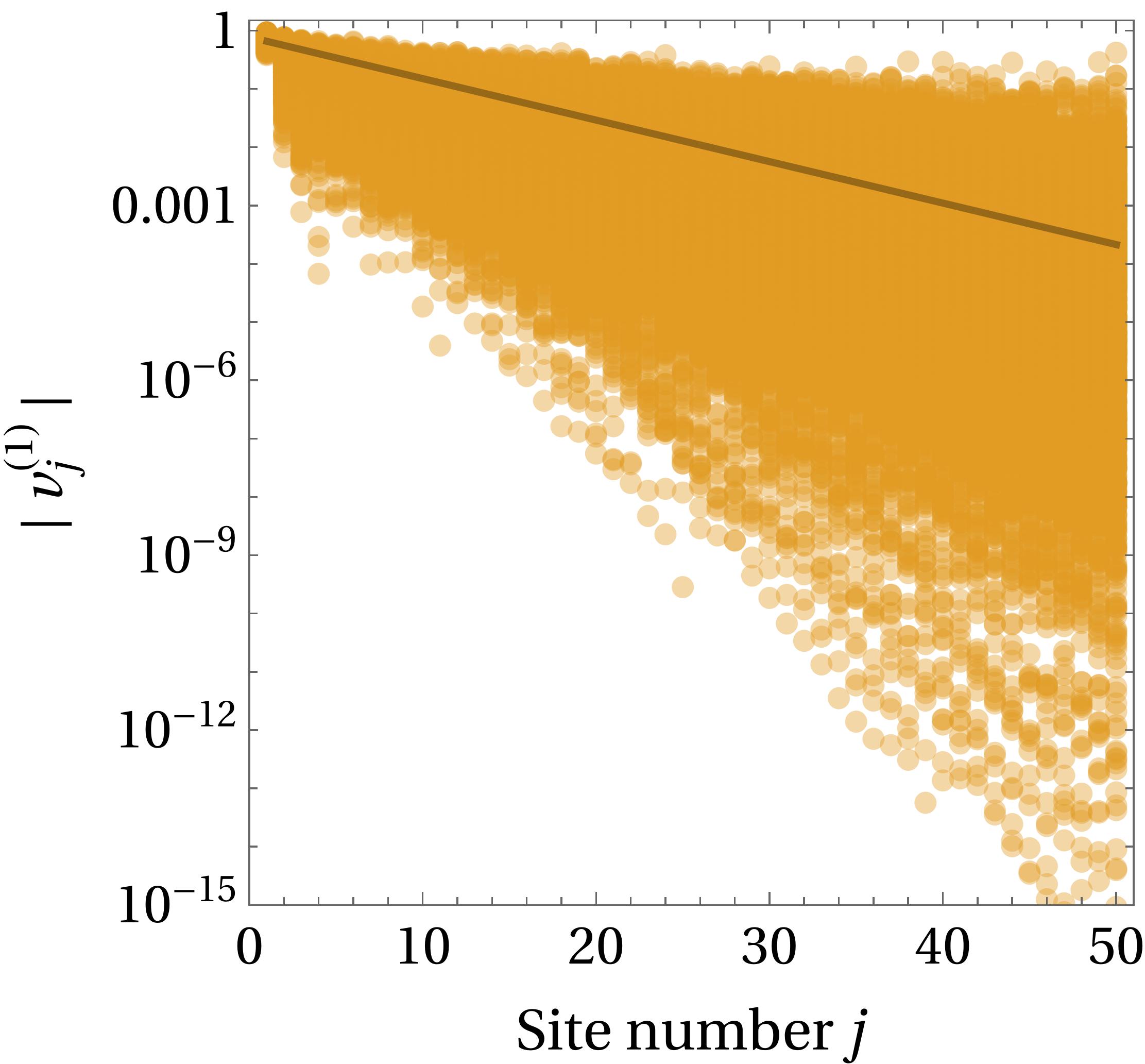}}
\caption{The magnitude of $j^{\rm th}$ element of an eigenvector $v^{(1)}$ of the matrix $M$ for $t/W = 0.1$ (left panel) and $t/W=0.3$ (right panel). The orange points represent values for $10^3$ samples of $M$, with $\epsilon_i$ drawn from a uniform distribution of random numbers ranging from 0 to $W=5$ TeV. The solid line represents an analytical expectation, eq. (\ref{vk_exp}), with the localisation length $L_1$ estimated using the ensemble average of a moving average, eq. (\ref{L_k}).}
\label{fig2}
\end{figure}

To demonstrate the localisation effect, we consider $N=50$ and randomly draw $\epsilon_i$ from a uniform random distribution over the interval $[0, W]$. Upon diagonalisation, we find the eigenvector whose largest magnitude element is located at the first site and call it $v^{(1)}$. The localisation length, $L_1$ in this case, is then estimated using eq. (\ref{L_k}). Taking a large ensemble of $M$, we display the magnitude of the components of $v^{(1)}$ at the different sites in Fig. \ref{fig2} for some sample values of $t/W$. We also show the result of an analytical expression, eq. (\ref{vk_exp}), using the ensemble-averaged value of the localisation length.
The localisation of the eigenvectors is stronger for small $t/W$ as can be seen from Fig. \ref{fig2}.

The localisation of the eigenvectors can be used to generate exponentially small couplings in QFT, as discussed in \cite{Craig:2017ppp}. For example, consider a pair of chiral fields, $f_{L,R}$, which couple to the chain only at the opposite ends. The case can be described by the following modification to eq. (\ref{LFC}):
\be \label{LFC_ex}
-{\cal L} \supset \mu\,\overline{f}_L R_1\,+\,\mu^\prime\,\overline{L}_N f_R\,+\sum_{i,j}\,M_{ij}\,\overline{L}_i R_j\,+\,{\rm h.c.}\,.\ee
In the physical bases, obtained by $R \to U R$ and $L \to U^* L$, this leads to
\be \label{LFC_ex1}
-{\cal L} \supset \sum_{i}\, \mu\, U_{1i}\,\overline{f}_L R_i\,+\,\sum_{i}\, \mu^\prime\, U^*_{Ni}\,\overline{L}_i f_R\,+\,\sum_{i}\, m_{i}\,\overline{L}_i R_i\,+\,{\rm h.c.}\,.\ee
For $\mu,\mu^\prime < m_i$, integrating out the fermions in the chain, one finds
\be \label{LFC_ex3}
-{\cal L}_{\rm eff} \supset \sum_{k=1}^N\,\frac{\mu \mu^\prime}{m_k} U_{1k} U^*_{Nk}\,\overline{f}_L f_R\,+\,{\rm h.c.}\, \equiv m_{\rm eff}\,\overline{f}_L f_R\,+\,{\rm h.c.}\,,\ee
with the effective mass 
\be \label{}
|m_{\rm eff}| = \sum_{k=1}^N\,\frac{\mu \mu^\prime}{m_k} \left|v^{(k)}_{1}\right| \left|v^{(k)}_{N}\right| \simeq  \sum_{k=1}^N\,\frac{\mu \mu^\prime}{m_k}\, \left|v^{(k)}_{k_0} \right|^2\,e^{-\frac{N-1}{L_k}}\,,\ee
where we have used the localisation property of the eigenvectors described by eq. (\ref{vk_exp}). For sufficiently large $N$, the effective mass of $f_{L,R}$ can be made exponentially suppressed with respect to the masses of the other fermions. A similar arrangement can be made with scalars or vector bosons.

It is noteworthy that while the fermion attached to the 1D lattice chain obtains exponentially suppressed mass, all the fermions in the chain have masses of the ${\cal O}(W)$. There are no exactly massless states in this setup. The separation of the two mass scales always requires some minimum $N$ depending on the size of $t/W$. If the disorder in the lattice is turned into a very particular order, as it happens in the models based on the clockwork mechanism \cite{Giudice:2016yja}, a localised massless state per chain can be obtained. However, the massless modes can also be arranged without relying on any elaborate ordering, as we show next.

\section{Massless modes from locality}
\label{sec:massless}
\begin{figure}[t!]
\centering
\subfigure{\includegraphics[width=0.85\textwidth]{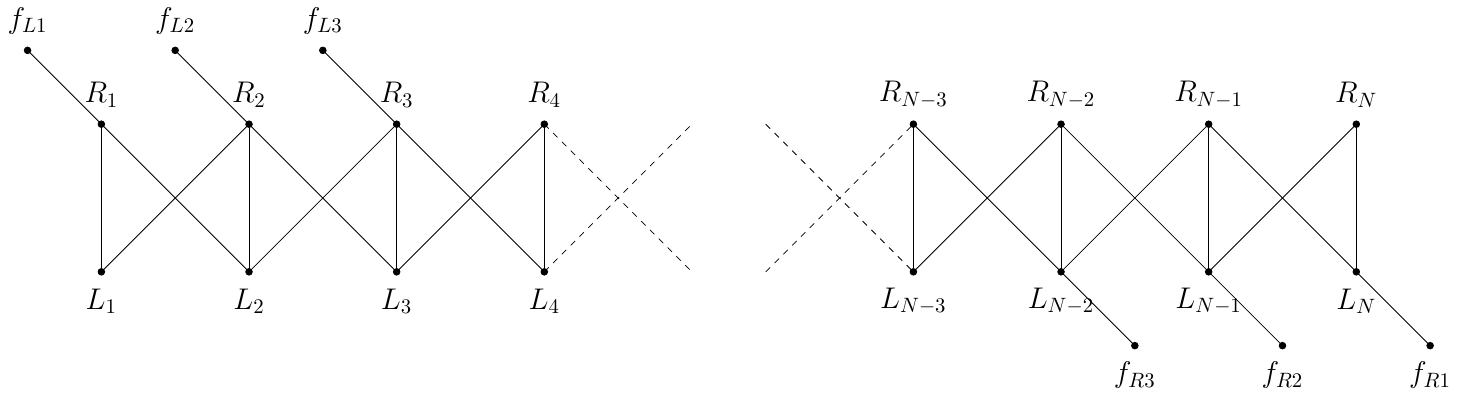}}
\caption{An example of chiral fermions' attachment to the Anderson chain leading to massless modes.}
\label{fig3}
\end{figure}
Consider a straightforward generalisation of the example discussed in the previous section for more than one copy of $f_{L,R}$. On the $N$-sites chain of Dirac fermions, the $N_f$ ($\leq N$) copies of chiral fermions, $f_{L \alpha}$ and $f_{R \alpha}$, are attached at the opposite ends, see Fig. \ref{fig3} for illustration. The interaction terms can be written as 
\beqa \label{L_fh}
-{\cal L} &\supset & \sum_{\alpha=1}^{N_f}\, \left(\mu_\alpha\,\overline{f}_{L \alpha} R_\alpha\,+\,\mu_\alpha^\prime\,\overline{L}_{(N-\alpha+1)} f_{R \alpha} \right)+ \sum_{i,j=1}^N\, M_{ij}\,\overline{L}_i R_j\,+\,{\rm h.c.}\,\nonumber \\
& \equiv & \overline{\cal F}_L\,{\cal M}\,{\cal F}_R\,+\,{\rm h.c.}\,, \eeqa
where we combine the $N_f$ flavours and $N$ site-fermions into,
\beqa \label{F_LR}
{\cal F}_L &=&(f_{L1},f_{L2},...,f_{L N_f},L_1,L_2,...,L_N)^T\,,\nonumber \\
{\cal F}_R &=&(f_{R1},f_{R2},...,f_{R N_f},R_1,R_2,...,R_N)^T\,.\eeqa
The square mass matrix ${\cal M}$ of dimension $(N_f+N)$ takes the form,
\be \label{M_fh}
{\cal M} = \left(\ba{cc} \left(0\right)_{N_f \times N_f} & \left(\mu_L \right)_{N_f \times N} \\ \left(\mu_R\right)_{N\times N_f}  & \left(M\right)_{N \times N}\ea\right)\,,\ee
where the subscript denotes the dimension of the respective block matrix. The latter can be expressed in more explicit form as,
\be \label{muLR_fh}
\mu_L = \begin{pmatrix}
(\mu_\alpha\,\delta_{\alpha \beta})_{N_f \times N_f} & \left(0 \right)_{N_f \times (N-N_f)}
\end{pmatrix}\,,~~\mu_R = \begin{pmatrix}
(0)_{(N-N_f) \times N_f} \\ \left(\mu^\prime_\alpha\, \delta_{\alpha,N-\beta+1}\right)_{N_f \times N_f} \end{pmatrix}\,. \ee
The matrix $M$ is in its usual form, given by eq. (\ref{M_FC}). 

It is noteworthy that the rank of the matrix ${\cal M}$ is always less than its dimension when $N>N_f$. Explicitly, we find
\be \label{rankM}
{\rm Rank}({\cal M}) = \left\{ \begin{array}{ll}
  2N_f & {\rm for}~N_f \leq  N < (2N_f-1)\,, \\
  N+1 & {\rm for}~ N \geq (2N_f-1)\,. \\
\end{array} \right.\ee
Consequently, only one pair of chiral fermions obtains their mass through the interactions in eq. (\ref{L_fh}) for $N \geq (2N_f-1)$, while $(N_f-1)$ of them remain exactly massless. It is the local structure of the mass matrix $M$ which gives rise to this rank deficit, as we explain below.

Upon integrating out the $N$-pairs of fermions for $\mu_L, \mu_R < M$, the effective mass matrix for $N_f$ flavours takes the form,
\be \label{M_eff}
m_{\rm eff} \simeq - \mu_L\, M^{-1}\,\mu_R\,.\ee
Using the form of $\mu_{L,R}$ given in eq. (\ref{muLR_fh}) and with the help of eq. (\ref{UMU}), the components of $m_{\rm eff} $ can be explicitly expressed as
\be \label{M_eff_ab}
\left(m_{\rm eff} \right)_{\alpha \beta} \simeq - \mu_\alpha\, \mu_\beta^\prime\, \sum_{k=1}^N \frac{1}{m_k}\, v_\alpha^{(k)}\, v_{N+1-\beta}^{(k)}\,.\ee
If $v^{(k)}$ and $m_k$ are the eigenvectors and eigenvalues of some general $N \times N$ matrix, then the above matrix has rank $N_f$. However, they have specific correlations in the present case due to the theory-space locality in  $M$.

Writing the eigenvalue equation for $M$,
\be \label{ev_eqn}
M\,v^{(k)} = m_k\,v^{(k)}\,,\ee
and substituting the tree-level form of $M$ from eq. (\ref{M_FC}), we find the following recursive relation between the different elements of an eigenvector:
\be \label{vk_recursive}
v^{(k)}_{i+1} + v^{(k)}_{i-1} + \left(\frac{\epsilon_i - m_k}{t}\right) v^{(k)}_i = 0\,.\ee
The above allows one to determine the entire eigenvector in terms of the parameters of $M$ and the eigenvalues $m_k$ through iterations. Using the above recursion relation, it can be shown that
\be \label{rank_meff} 
{\rm Rank}(m_{\rm eff})=1\,~~{\rm for}~N \geq (2N_f-1)\,. \ee
The proof is given in Appendix \ref{app:rank}. Therefore, the matrix $m_{\rm eff}$ leads to $(N_f-1)$ massless modes.

The mass of the only massive state provided by $m_{\rm eff}$ is given by
\beqa \label{Tr}
{\rm Tr}(m_{\rm eff}) &=& -\sum_{\alpha=1}^{N_f}\,\sum_{k=1}^{N}\,\frac{\mu_\alpha \mu^\prime_\alpha}{m_k}\,v_\alpha^{(k)}\,v_{N+1-\alpha}^{(k)}\, \nonumber \\
&\simeq &  -\sum_{\alpha,k}\,\frac{\mu_\alpha \mu^\prime_\alpha}{m_k}\,\left(v^{(k)}_{k_0}\right)^2\,\exp\left[-\frac{N+1-2\alpha}{L_k} \right]\, \nonumber \\
&\sim & {\cal O}\left(\frac{\mu_{N_f} \mu^\prime_{N_f}}{m_k}\right) \exp \left[-\frac{N-(2N_f-1)}{L_k} \right]\,. \eeqa
Here, we have used the localisation properties of the eigenvectors, eq. (\ref{vk_exp}).  Consequently, the mass is exponentially suppressed compared to the scale ${\cal O}(\mu \mu^\prime/W)$ for $N \gg (2N_f -1)$.

The eigenvectors corresponding to the $N_f-1$ massless modes, namely $\Omega_{L,R}^{(p)}$ with $p=1,\,...,\,N_f-1$, can be obtained from eq.~(\ref{M_fh}) with the help of eq.~(\ref{ev_eqn}). The explicit computation is provided in Appendix~\ref{app:massless_comp}. We find them in the following form:
\beqa \label{Omega_LR}
\Omega_L^{(p)} &=& \left(u^{(p)}_{L,1},\,u^{(p)}_{L,2},\,...,\,u^{(p)}_{L,{N_f}},\,w^{(p)}_{L,1},\,w^{(p)}_{L,2},\,...,\,w^{(p)}_{L,N_f-1},0,0,\,...,\,0 \right)\,,\nonumber \\
\Omega_R^{(p)} &=& \left(u^{(p)}_{R,1},\,u^{(p)}_{R,2},\,...,\,u^{(p)}_{R,{N_f}},0,0,\,...,\,0,\,w^{(p)}_{R,{N+2-N_f}},\,w^{(p)}_{R,{N+3-N_f}},\,...,\,w^{(p)}_{R,N} \right)\,.
\eeqa
Notably, the massless modes are linear combinations of only $2N_f-1$ of the total $N_f+N$ fields of a given chirality. Specifically, they involve $f_{L1},\,...,\,f_{L{N_f}},\,L_1,\,...,\,L_{N_f-1}$ in the case of left-chiral fields, and $f_{R1},\,...,\,f_{R{N_f}},\,R_{N+2-N_f},\,...,\,R_{N}$ in the case of right-chiral fields. The other $N_f$ linear combinations of these $2N_f-1$ fields acquire non-vanishing masses. The values of the non-zero elements $u_{L(R)}$ and $w_{L(R)}$ in eq.~(\ref{Omega_LR}) depend on the parameters $\mu_{L(R)}$, $\epsilon_i$, and $t$, as outlined in Appendix~\ref{app:massless_comp}. For generic values of these parameters, the elements are found numerically to take non-hierarchical values.

The existence of $N_f-1$ massless modes implies a global $U(N_f-1)_L \times U(N_f-1)_R$ symmetry for the terms in eq.~(\ref{L_fh}). In the physical basis, its action is to rotate non-trivially the $N_f-1$ massless modes as
\be \label{sym_rot1}
f^\prime_{L,p} \to \sum_{q=1}^{N_f-1}\,\left(U_L\right)_{pq}\,f^\prime_{L,q}\,,~~f^\prime_{R,p} \to \sum_{q=1}^{N_f-1}\,\left(U_R\right)_{pq}\,f^\prime_{R,q}\,,
\ee
where $U_{L,R}$ are $(N_f-1)$-dimensional unitary matrices representing arbitrary rotations. Since the massless modes consist of only $2N_f-1$ fields, the action of this symmetry on the original basis can be seen as rotating non-trivially only this subset of fields. Explicitly, it is given by
\be \label{sym_rot2}
{\cal F}_{L,m} \to \sum_{n}\,\left({\cal U}_L\right)_{mn}\,{\cal F}_{L,n}\,,~~{\cal F}_{R,m^\prime} \to \sum_{n^\prime}\,\left({\cal U}_R\right)_{m^\prime n^\prime}\,{\cal F}_{R,n^\prime}\,,
\ee
where $m,n=1,\,...,\,2N_f-1$, while $m^\prime, n^\prime=1,\,...,\,N_f,\,N+2-N_f,\,...,\,N$, and
\beqa \label{rotations}
\left({\cal U}_L\right)_{mn} &=& \delta_{mn}-\sum_{p,q=1}^{N_f-1}\,\Omega^{(p)}_{L,m}\,\Omega^{(q)}_{L,n}\,\left(\delta_{pq} - \left(U_L\right)_{pq}\right)\,, \nonumber \\
\left({\cal U}_R\right)_{m^\prime n^\prime} &=& \delta_{m^\prime n^\prime}-\sum_{p,q=1}^{N_f-1}\,\Omega^{(p)}_{R,m^\prime}\,\Omega^{(q)}_{R,n^\prime}\,\left(\delta_{pq} - \left(U_R\right)_{pq}\right)\,.
\eeqa
The remaining $N-N_f+1$ fields in ${\cal F}_{L,R}$ transform trivially under this symmetry.

Evidently, the existence of the $U(N_f-1)_L \times U(N_f-1)_R$ symmetry of ${\cal M}$, leading to the $N_f-1$ massless modes, can be attributed to the topology of the interactions between the chiral and lattice fermions, and it results from the strict locality in the lattice interactions. It is, therefore, expected that the non-local interactions make the massless modes massive in general. If the non-local couplings among the chain fermions are strong enough and disordered independently of the local ones, they can lead to complete delocalisation of the mass eigenstates, destroying the main features of the underlying mechanism. On the other hand, if such effects are small and ordered, one expects only a small correction to the tree-level arrangement.

\section{Non-local effects in Abelian gauge theories}
\label{sec:nonlocal}
Let us consider a case in which the interactions between the lattice fermions beyond the nearest neighbour arise from quantum effects. This is quite natural to expect if the fermions in the chain have additional interactions with massive gauge bosons or scalars, which radiatively induce such effects. For calculability and simplicity, we consider gauge interactions, which are only abelian. To maintain generality within this choice, we assume that the gauge group $G$ under which the chain fermions are charged has $N_G$ number of $U(1)$ factors,
\be \label{G}
G = \prod_{a=1}^{N_G}\,U(1)_a \,.\ee
The most general gauge interactions involving fermions in the chain under the aforementioned assumptions can be written as
\be \label{Lg}
-{\cal L}_{\rm gauge} = \sum_{i=1}^N \sum_{a=1}^{N_G}\,g\,\left(q^{(a)}_{L i}\,\overline{L}_i \gamma^\mu L_i + q^{(a)}_{R i}\,\overline{R}_i \gamma^\mu R_i\right) X_\mu^{(a)}\,, \ee
where $q_{Li}^{(a)}$ is a charge of $L_i$ under the $U(1)_a$ and so on. $X_\mu^{(a)}$ is the gauge boson of  $U(1)_a$. It is straightforward to see that the choice $N_G=N$ and $q_{L i}^{(a)}=q_{R i}^{(a)} \propto \delta_{ia}$ corresponds to the gauged version of ``quivers" that can be used to obtain the desired structure of $M$ as discussed in the previous section.

In the physical basis of the fermions and gauge bosons, obtained in general by rotations $L \to U_L L$, $R \to U_R R$ and $X_\mu \to {\cal R} X_\mu$, eq. (\ref{Lg}) can be recast into 
\be \label{Lg2}
-{\cal L}_{\rm gauge} = \sum_{i,j=1}^N\, \sum_{a,b=1}^{N_G}\,g \left(\left(Q^{(a)}_L\right)_{ij}\,\overline{L}_i \gamma^\mu L_j + \left(Q^{(a)}_R\right)_{ij}\,\overline{R}_i \gamma^\mu R_j\right) {\cal R}_{ab} X_\mu^{(b)}\,. \ee
Here,
\be \label{Q}
Q^{(a)}_{L,R} = U_{L,R}^\dagger\,q^{(a)}_{L,R}\,U_{L,R}\,,\ee
are $N \times N$ matrices which need not necessarily be diagonal. ${\cal R}$ is $N_G \times N_G$ real orthogonal matrix, and it is non-trivial when the different gauge bosons mix.

In general, the above gauge interactions can contribute to $M$ through radiative corrections and can modify the tree-level structure given in eq. (\ref{M_FC}). We compute such corrections at the leading order under the assumption that the abelian symmetries $G$ are completely broken at a scale around $M$.  At 1-loop, the mass terms of fermions in the Anderson chain can receive corrections from diagrams involving them and the gauge bosons of $G$ in the loop. Such correction has been explicitly computed in \cite{Mohanta:2022seo} for a different class of models, and we closely follow the notations and methods with appropriate modification and generalisation.

The 1-loop self-energy correction in Feynman-'t Hooft gauge at vanishing external momentum results into 
\be \label{Sigma}
\Sigma_{ij}(p=0) = (\sigma_L)_{ij}\,P_L\,+\,(\sigma_R)_{ij}\,P_R\,,\ee
where $P_{L,R}$ are left- and right-chiral projection matrices and
\be \label{sigma_LR}
\left(\sigma_{L,R}\right)_{ij} = \frac{g^2}{4 \pi^2}\,\sum_{a,b,c} {\cal R}_{ac} {\cal R}_{bc}\,\sum_{k=1}^N \left(Q^{(a)}_{R,L}\right)_{ik} \left(Q^{(b)}_{L,R}\right)_{kj}\,m_k\,B_0[M_c^2,m_k^2]\,.\ee
Here, $m_k$ is the tree-level mass of the $k^{\rm th}$ fermion as earlier and $M_c$ is mass of $X^{(c)}_\mu$. The function 
\be \label{B0}
B_0[m_A^2,m_B^2] = \Delta_\epsilon + 1 - \frac{m_A^2 \ln \frac{m_A^2}{\mu_R^2} - m_B^2 \ln \frac{m_B^2}{\mu_R^2}}{m_A^2 - m_B^2}\,,\ee
is a loop integration function evaluated in the dimensional regularisation scheme at the renormalisation scale $\mu_R$ \cite{Passarino:1978jh}, and 
\be \label{Delta_ep}
\Delta_\epsilon = \frac{2}{\epsilon} - \gamma_E + \ln 4\pi\,.\ee
The shift from the tree-level mass matrix generated by the above corrections can be evaluated as
\be \label{dM0}
\delta M = U_L\,\sigma_R\,U_R^\dagger\,, \ee
which, after some straightforward simplification, can be expressed as
\be \label{dM}
\delta M_{ij} = \frac{g^2}{4 \pi^2}\,\sum_{a,b,c} {\cal R}_{ac} {\cal R}_{bc}\,q^{(a)}_{L i} q^{(b)}_{R j}\sum_{k}\, (U_L)_{ik}\,(U_R^\dagger)_{kj}\,m_k\,B_0[M_c^2,m_k^2]\,.\ee
Note that the above $\delta M$ is evaluated for the general tree-level mass matrix, and the above expression is valid even if the latter is not in any specific form.

Consider now the case of $M$ given by eq. (\ref{M_FC}). Since $U_R=U^*_L=U$ in this case, the radiative shift can be simplified to
\be \label{dM1}
\delta M_{ij} = \frac{g^2}{4 \pi^2}\,\sum_{a,b,c} {\cal R}_{ac} {\cal R}_{bc}\,q^{(a)}_{L i} q^{(b)}_{R j}\,\sum_{k}\, U^*_{ik}\,U^*_{jk}\,m_k\,B_0[M_c^2,m_k^2]\,.\ee
The divergent part of $\delta M$, i.e. the coefficient of $\Delta_\epsilon$, can be further simplified to
\be \label{Div}
{\rm Div.}\left[\delta M_{ij} \right] \propto \sum_a q^{(a)}_{L i} q^{(a)}_{R j}\,\sum_k U^*_{ik}\,U^*_{jk}\,m_k = \sum_a q^{(a)}_{L i} q^{(a)}_{R j}\,M_{ij}\,. \ee
Here, we have used the orthogonality of ${\cal R}$ and eq. (\ref{UMU}). As a result, the divergent part of $\delta M_{ij}$ vanishes if the corresponding $M_{ij}=0$. Therefore, the radiative corrections to the non-nearest-neighbour terms in $M$ are finite and calculable, as expected in a renormalizable theory.

The finite corrections to non-local terms in $M$ are in-principle non-vanishing and depend on the gauge charges and the tree-level masses of the fermions in the chain. As discussed before, the non-local effects can affect the localisation property of the eigenvectors as well as induce masses for the otherwise massless modes. However, such effects arise through eq. (\ref{dM1}) are not completely chaotic and uncontrolled in their magnitude. The disorder in these elements is highly correlated with that in the tree-level mass matrix. For explicit understanding, we consider some simple examples of $G$ in the next section and quantitatively evaluate these effects.

\section{Some examples}
\label{sec:examples}
In all the following examples, the fermions in the chain are considered to be vectorlike under $G$, i.e. $q^{(a)}_{Li}=q^{(a)}_{R i} \equiv q^{(a)}_i$ for each $i$. The choice is sufficient to make the underlying gauge symmetry free from anomaly. With this simplification, we consider four examples of $G$ as the following.

\subsection{Site universal $U(1)$}
\label{subsec:universal}
The simplest non-trivial case corresponds to a single abelian gauge symmetry $G=U(1)$ under which all the fermions are universally charged, i.e. $q_i^{(a)} = q$. In this case, the 1-loop correction to tree-level mass matrix, eq. (\ref{dM1}), can be further simplified to
\be \label{dM1_uni}
\delta M_{ij} = \frac{g^2q^2}{4 \pi^2}\,\sum_{k}\, U^*_{ik}\,U^*_{jk}\,m_k\,B_0[M_X^2,m_k^2]\,,\ee
where $M_X$ now corresponds to the mass of the gauge boson of the broken $U(1)$. 
The non-nearest-neighbour elements of $\delta M_{ij}$ are non-vanishing, which gives rise to loop-suppressed and calculable non-local effects in the lattice chain.

Using eq. (\ref{UMU}), it is straightforward to verify that $M$ and $\delta M$ commute with each other in this case. Therefore, the one-loop corrected mass matrix, $M+\delta M$, is also diagonalised by the same unitary matrix $U$. Consequently, the localisation of the eigenvectors remains the same even after the corrections, despite the generation of non-local terms in the fermion mass matrix. This can be attributed to the fact that the induced corrections are strongly correlated with the diagonal disorder. Therefore, the case of site universal abelian gauge interaction provides a unique example in which the non-local couplings in the Anderson chain are induced without any change in the localisation properties of the eigenvectors. Some phenomenological advantages of this are outlined in the next section.

\begin{figure}[t!]
\centering
\subfigure{\includegraphics[width=0.42\textwidth]{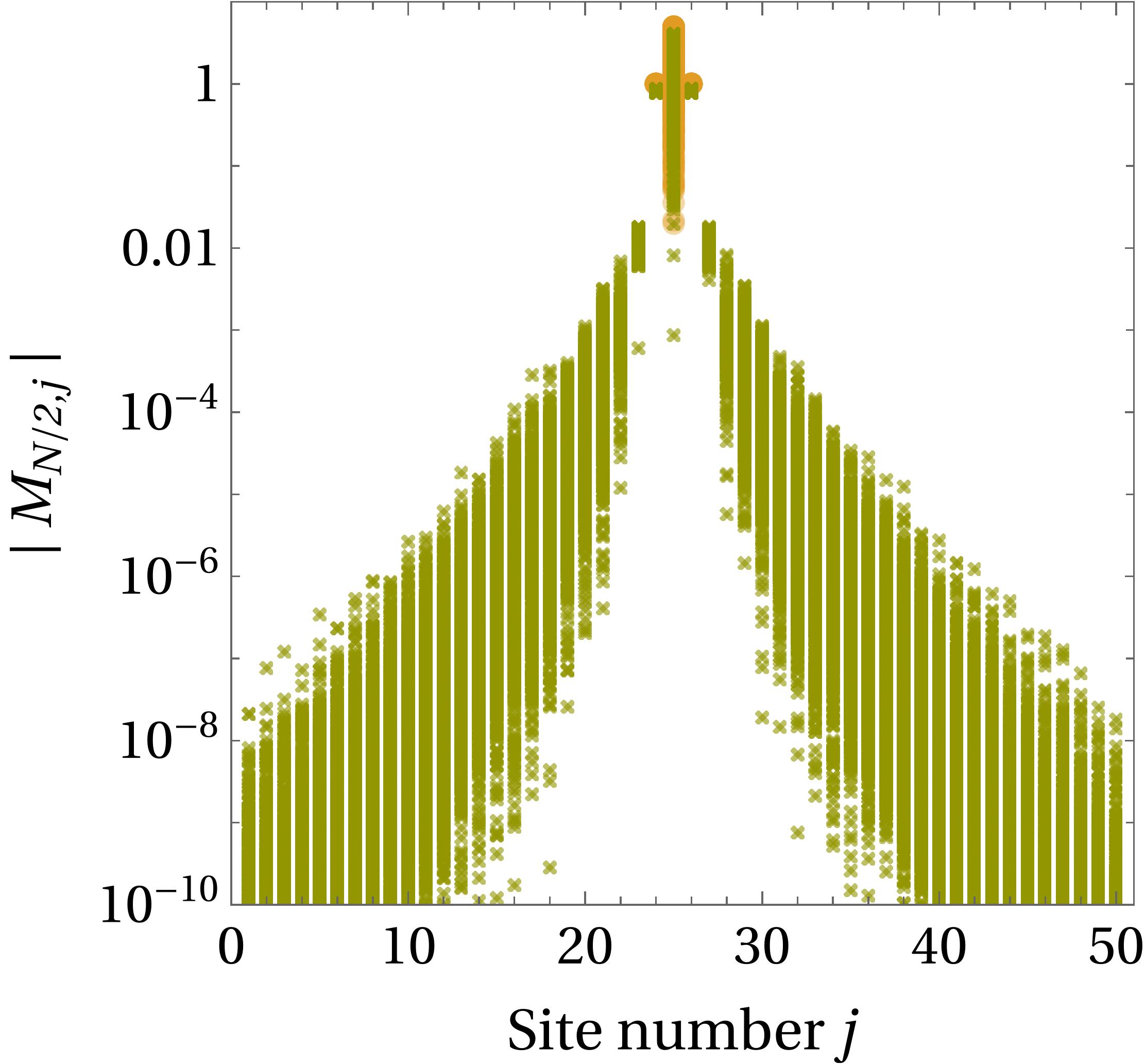}}\hspace*{0.5cm}
\subfigure{\includegraphics[width=0.43\textwidth]{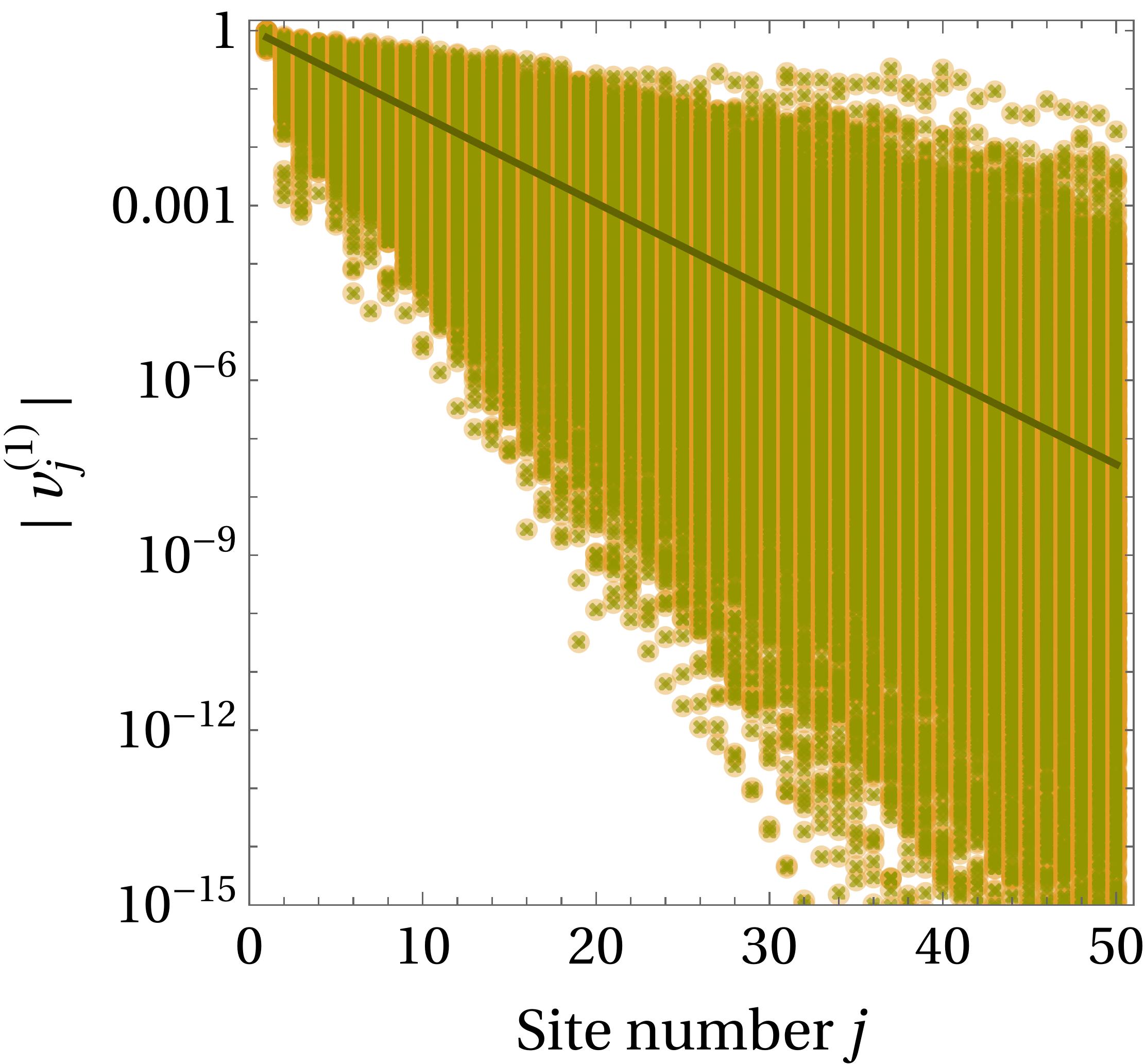}}
\caption{Magnitudes of $M_{N/2,j}$ (left panel) and $v^{(1)}_j$ (right panel) with respect to the site number $j$ for $10^3$ samples and $t/W=0.2$. The values of $\epsilon_i$ are drawn from a uniform distribution of random numbers over the interval $[0,W=5]$ TeV. The orange dots are the values corresponding to tree-level mass matrices, while the green crosses denote the values after the 1-loop corrections are switched on. The dark orange and green solid lines in the right panel represent eq. (\ref{vk_exp}) with the localisation lengths computed using the ensemble average of eq. (\ref{L_k}) for the tree-level and 1-loop corrected $M$, respectively. The corrections are evaluated for site universal $U(1)$ gauged interactions, as discussed in the text.}
\label{fig4}
\end{figure}

To understand the nature and magnitude of the non-local terms induced by quantum correction, we compute $\delta M$ for $t/W=0.2$, $N=50$ and set $gq=1$, $W=5$ TeV and $M_X = 0.2\, W$. The renormalisation scale $\mu_R$ is taken to be the $Z$-boson mass scale, $M_Z$. Replacing the tree-level matrix by its radiatively corrected version, $M \to M + \delta M$, we identify the specific eigenvector $v^{(1)}$ as discussed previously in section \ref{sec:anderson}. For the demonstration of the strength of non-local effects, we choose the central row of the fermion mass matrix as a reference. The magnitude of elements of this row and localisation in $v^{(1)}$ are shown in Fig. \ref{fig4} with and without the 1-loop corrections. 

It can be seen from the left panel in Fig. \ref{fig4} that the non-local couplings in $M$ generated at 1-loop are non-vanishing and hierarchical. The latter feature is essentially due to the Anderson localisation through the tree-level mass itself. Using eq. (\ref{vk_exp}) in (\ref{dM1_uni}), the size of such terms can be estimated as
\beqa \label{dM1_universal_size}
\delta M_{ij} &\sim& \frac{g^2q^2}{4 \pi^2} \left[M_{ij} \left(1-\ln \frac{M_X^2}{\mu_R^2}\right)+\sum_{k} m_k\, e^{-\frac{|i-k_0|+|j-k_0|}{L_k}} \left(\frac{m_k^2}{M_X^2} \ln\frac{m_k^2}{M_X^2} + {\cal O}\left(\frac{m_k^4}{M_X^4}\right)\right)\right] \eeqa
for $m_k \ll M_X$, and 
\beqa \label{dM1_universal_size_2}
\delta M_{ij} &\sim& \frac{g^2q^2}{4 \pi^2} \left[M_{ij} \left(1-\ln \frac{M_X^2}{\mu_R^2}\right)-\sum_{k} m_k\, e^{-\frac{|i-k_0|+|j-k_0|}{L_k}} \left(\ln\frac{m_k^2}{M_X^2} + {\cal O}\left(\frac{M_X^2}{m_k^2}\right)\right)\right]\,,\eeqa
for $m_k \gg M_X$. Consequently, the non-vanishing $\delta M_{ij}$ induced through quantum corrections is largest for $i=j=k_0$ and decreases exponentially away from this point. The incorporation of exponentially decaying off-site terms in general, and their impact on the localization of the eigenstates, is studied in \cite{Tropper:2020yew}. In the present case, such hierarchical non-local terms naturally arise as an effect of quantum corrections. The right panel of Fig. \ref{fig4} shows that the localization properties of the eigenvectors after the corrections remain unchanged, as already anticipated in this case.

\subsection{Odd-even $U(1)$}
Consider an abelian gauge interaction, which is site-dependent. A simple example is a single $U(1)$ with the fermions at the odd-even sites having charges $q^{(a)}_i=\pm q$, respectively. In this case, eq. (\ref{dM1}) reduces to
\be \label{dM1_oddeven}
\delta M_{ij} = \frac{g^2q^2}{4 \pi^2}\,(-1)^{i+j}\,\sum_{k}\, U^*_{ik}\,U^*_{jk}\,m_k\,B_0[M_X^2,m_k^2]\,.\ee
Because of the site-dependent sign factor, $\delta M$ does not commute with $M$ in this case, and hence the 1-loop corrected mass matrix $M+\delta M$ has eigenvectors different from those of $M$.

\begin{figure}[t!]
\centering
\subfigure{\includegraphics[width=0.42\textwidth]{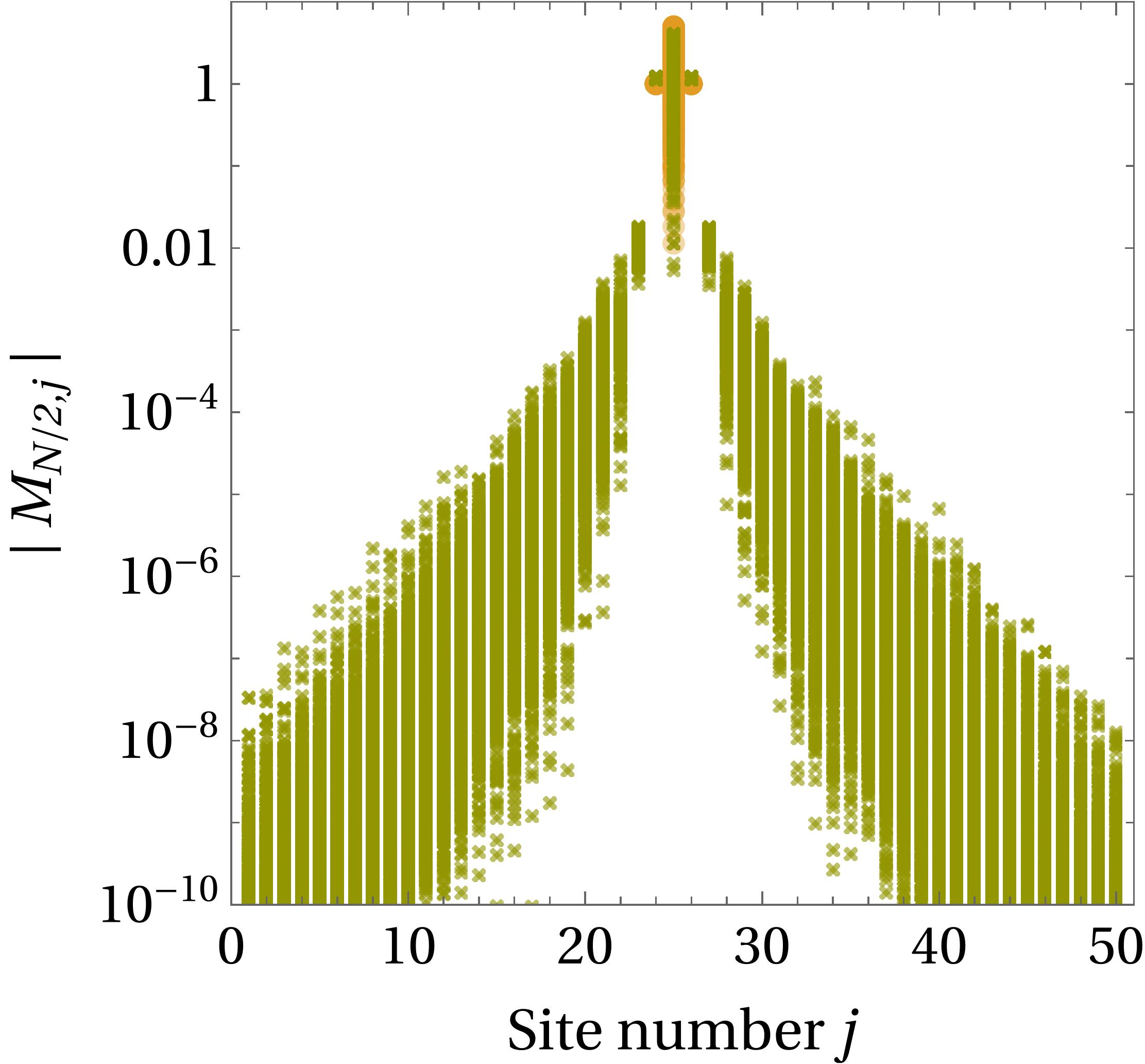}}\hspace*{0.5cm}
\subfigure{\includegraphics[width=0.43\textwidth]{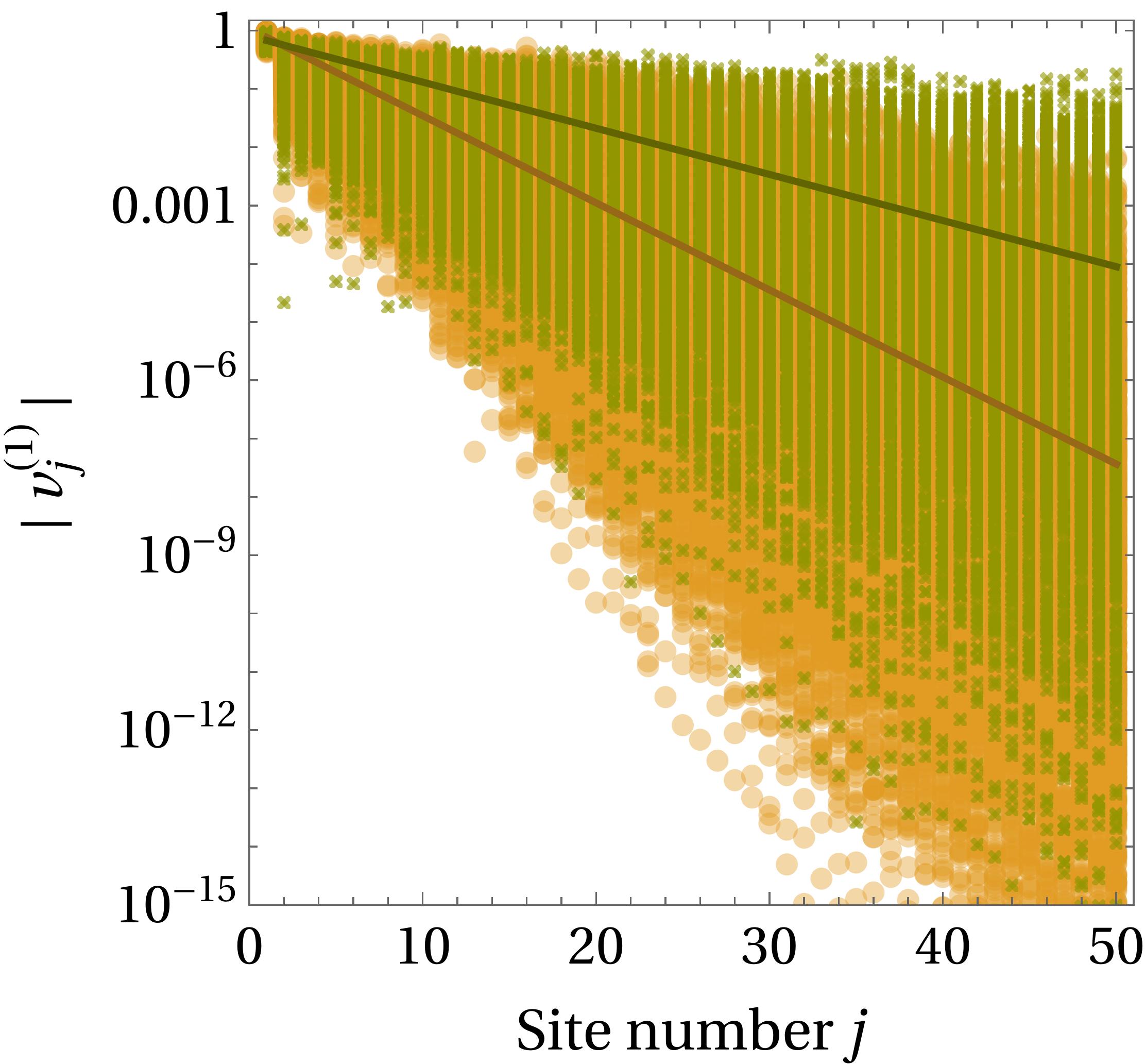}}
\caption{Same as Fig. \ref{fig4} but for the quantum corrections evaluated for site dependent odd-even $U(1)$.}
\label{fig5}
\end{figure}
The radiative corrections and shift in the localisation of the eigenvectors induced by them are displayed in Fig. \ref{fig5}. The loop effects lead to a larger localisation length, resulting in milder localisation of the eigenvectors in this case. As it can be seen, the average value of the $|v^{(1)}_{50}|$ increases by three to four orders of magnitude. The overall strength of the non-local couplings is similar to the ones seen in the previous case of site-universal $U(1)$, however, the effective width of the diagonal disorder increases resulting in relatively milder localisation in the eigenvector.

\subsection{$U(1)$ site-number}
Another example of site-dependent abelian gauge symmetry is the $U(1)$ site-number. The fermions at the $j^{\rm th}$ site are charged as $q^{(a)}_j=2 \pi j/N$. The charges are devised in such a way that the furthest fermions in the chain have the strongest gauge interactions, while the nearer ones are relatively less strongly interacting. The factor $2 \pi/N$ is introduced to ensure that the interactions remain perturbative. The gauge symmetry can also be used to arrange the tree-level ${\cal L}_{\rm FC}$ given in eq. (\ref{LFC}). The onsite interactions are gauge invariant, while $t$ can be thought of as a spurion with $U(1)$ charge $\pm 2 \pi/N$. The renormalizability along with an absence of scalars with other charges ensures that the tree-level mass matrix $M$ takes the desired form of eq. (\ref{M_FC}).

\begin{figure}[t!]
\centering
\subfigure{\includegraphics[width=0.42\textwidth]{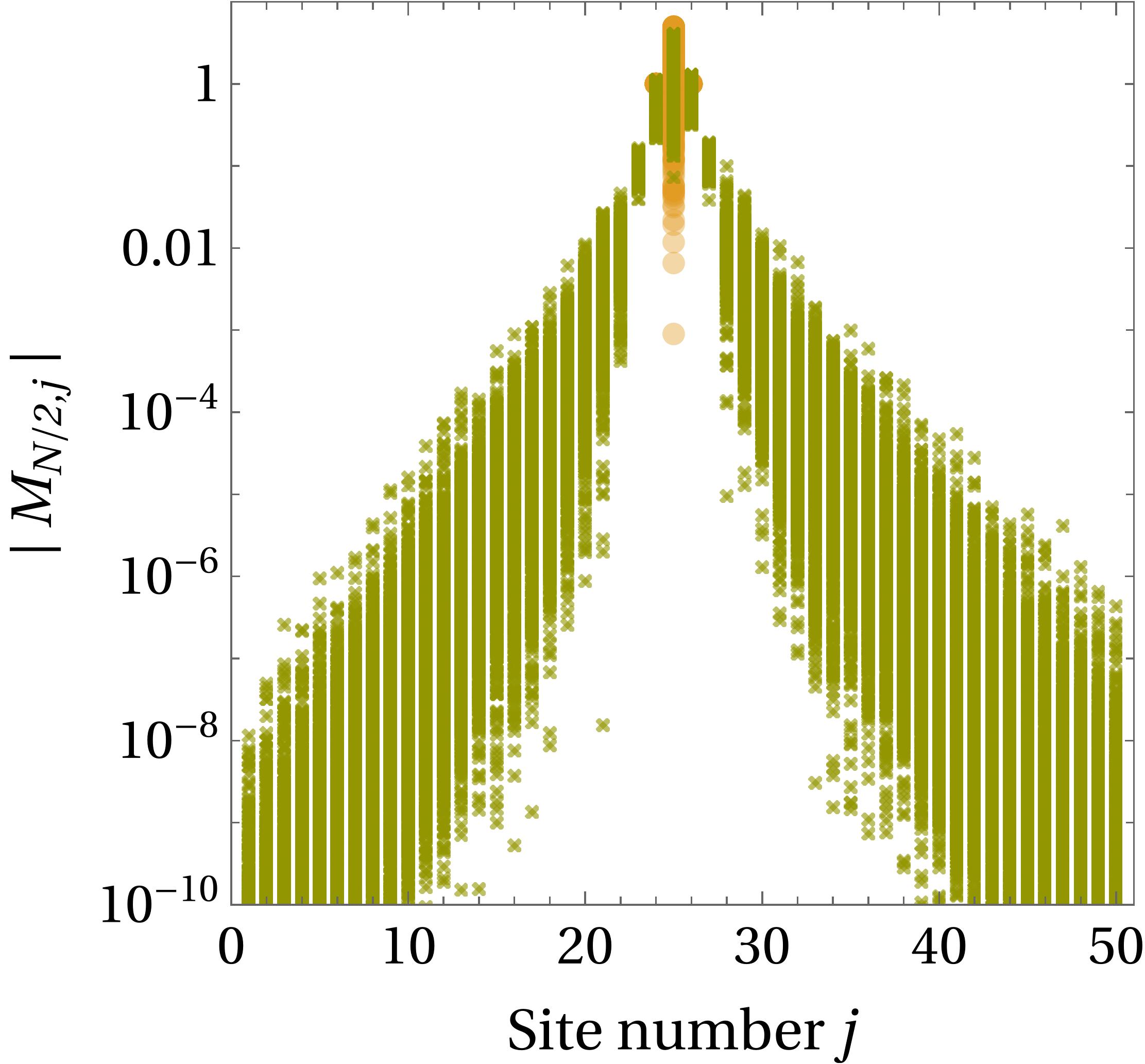}}\hspace*{0.5cm}
\subfigure{\includegraphics[width=0.43\textwidth]{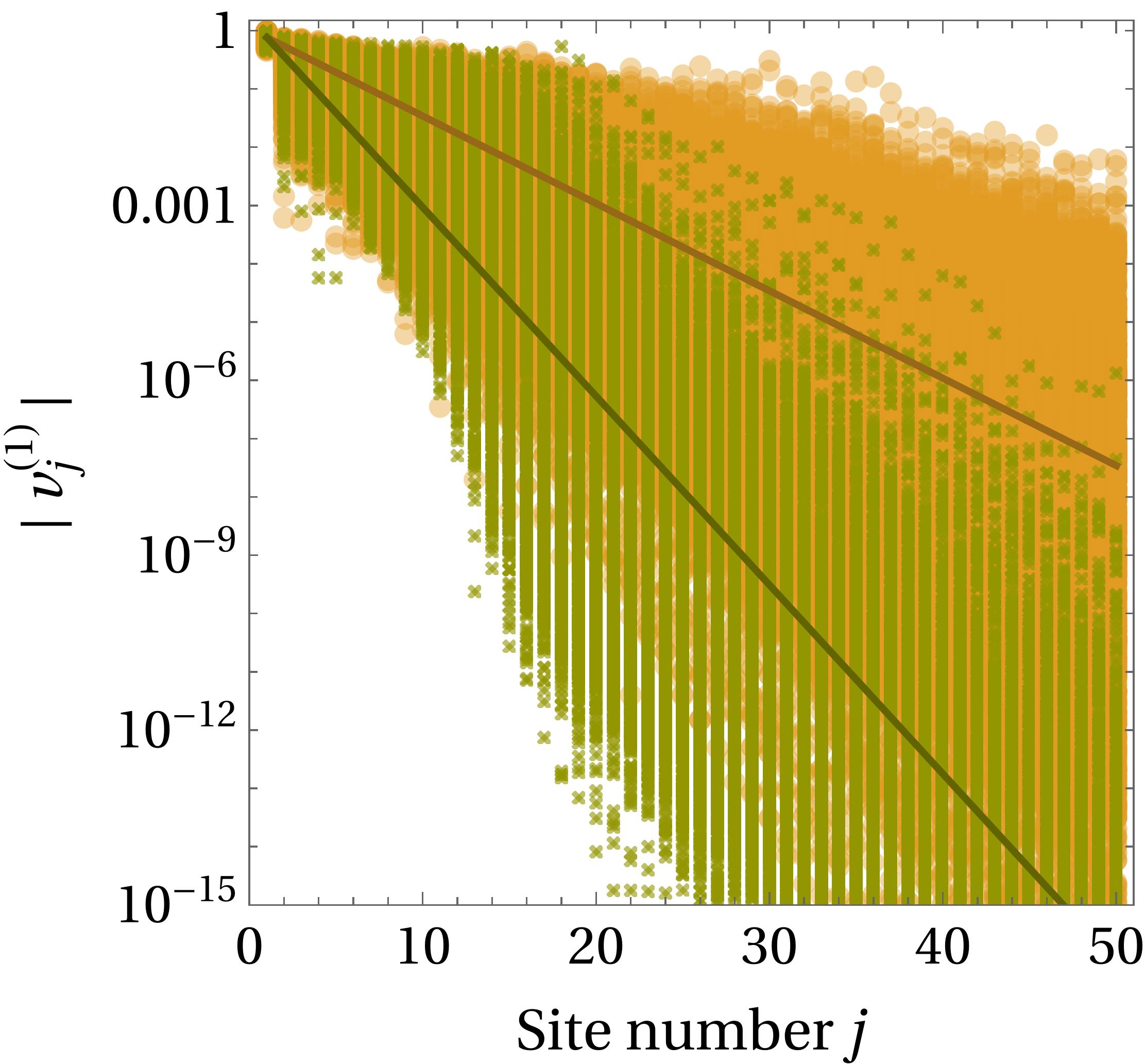}}
\caption{Same as Fig. \ref{fig4} but for the quantum corrections evaluated for $U(1)$ site-number.}
\label{fig6}
\end{figure}

The non-local couplings will be generated once the underlying $U(1)$ is broken. The corrections take the form,
\be \label{dM1_sitenum}
\delta M_{ij} = \frac{g^2}{N^2}\,i\, j\,\sum_{k}\, U^*_{ik}\,U^*_{jk}\,m_k\,B_0[M_X^2,m_k^2]\,.\ee
Again, $\delta M_{ij}$ are hierarchical due to exponential localisation of the eigenvectors despite site number entering as a weight factor. Their values and modification in the localisation of the eigenvectors, evaluated for the same choice of $t/W$ and other parameters as earlier, are shown in Fig. \ref{fig6}. 

As it can be noticed, the quantum corrections effectively decrease the width of the distribution from which the values of the diagonal parameters $\epsilon_i$ are randomly drawn. Moreover, the strength of the nearest-neighbour interactions also gets modified significantly through these corrections. As they become even smaller than the tree-level value of $t/W$, the localisation length becomes nearly half of the original one. This results in a rather strong localisation of the eigenvectors, as can be seen from the right panel of Fig. \ref{fig6}. For example, the desired suppression in the magnitude of $v^{(1)}_j$ can be achieved even at $j/2$ when the gauge interactions in the fermionic chain are enabled. The generation of non-local effects and stronger localisation of the eigenvectors can have important implications in the phenomenological applications.

\subsection{Gauged quivers}
A quiver of $N$ factors of $U(1)$ can also be used to get the desired form of tree-level $M$ as discussed in section \ref{sec:anderson}. If this symmetry is gauged and spontaneously broken, it can induce quantum correction in $M$. Unlike the previous examples, there are multiple gauge bosons here and the loop-corrections depend not only on their masses but also on the mixing among them. The latter in turn depends on the pattern of the breaking of $G=\prod_{a=1}^N U(1)_a$. In the following, we discuss a minimal scenario in which the scalars responsible for generating nearest-neighbour interactions at tree-level are the sole source of the breaking of $G$. The fermions located at the $j^{\rm th}$ site in the chain are charged under the $G$ with charges $q^{(a)}_j = \delta_{aj}$. The diagonal couplings in $M$ are, therefore, gauge invariant. The off-diagonal nearest-neighbour couplings can be arranged through a set of $N+1$ scalar fields, $\phi_j$ ($j=0,1,...,N$) with charges $\delta_{ja}-\delta_{j+1,a}$ under the $G$. The desired form of the tree-level $M$, eq. (\ref{M_FC}), is obtained when all the $\phi_j$ acquire universal vacuum expectation value (VEV).

The VEV of $\phi_j$ can also give rise to the gauge boson masses. The relevant terms arise from the kinetic terms of $\phi_j$ which are given by,
\be \label{Dphi}
D_\mu \phi_j= \partial_\mu \phi_j + i g\,  \sum_{a=1}^N\,  (\delta_{ja}-\delta_{j+1,a})\,  X_\mu^{(a)}\,\phi_j\,. \ee
This leads to
\beqa \label{Dphi2}
\sum_{j=0}^N\,|D_\mu \phi_j|^2 &\supset&  g^2\,\sum_{j=0}^N\,|\phi_j|^2\, \sum_{a,b}\,  (\delta_{ja}-\delta_{j+1,a}) (\delta_{jb}-\delta_{j+1,b})\,X^{(a)} X^{(b)}\, \nonumber \\
&=&g^2 \sum_{j=1}^N  \left(|\phi_j|^2+|\phi_{j-1}|^2\right) X^{(j)} X^{(j)} - 2g^2 \sum_{j=1}^{N-1}  |\phi_j|^2\,X^{(j)} X^{(j+1)}\,, \eeqa
where we have suppressed the Lorentz indices for brevity. Once the  $\phi_j$ takes non-vanishing values in the vacuum, the above terms generate the diagonal and nearest-neighbour off-diagonal elements in the gauge boson mass matrix.  Note that if the VEVs of $\phi_j$ are chosen randomly, the gauge boson mass matrix also turns out to be of the type of the Hamiltonian of the Anderson tight-binding model. For the analysis in the given case, we however assume universal VEVs: $\langle |\phi_1|^2 \rangle = \langle |\phi_2|^2 \rangle = ... =\langle |\phi_N|^2 \rangle \equiv v_\phi^2 $. The gauge-boson mass terms in the basis $\left({\cal M}^2_X\right)_{ab} X^{(a)} X^{(b)}$ then takes a very simple form,
\be \label{M_GB}
\left({\cal M}_X^2\right)_{ab} =g^2 v_\phi^2\, \left(2\delta_{ab} -   \delta_{a+1,b} - \delta_{a,b+1}\right)\,.\ee

\begin{figure}[t!]
\centering
\subfigure{\includegraphics[width=0.42\textwidth]{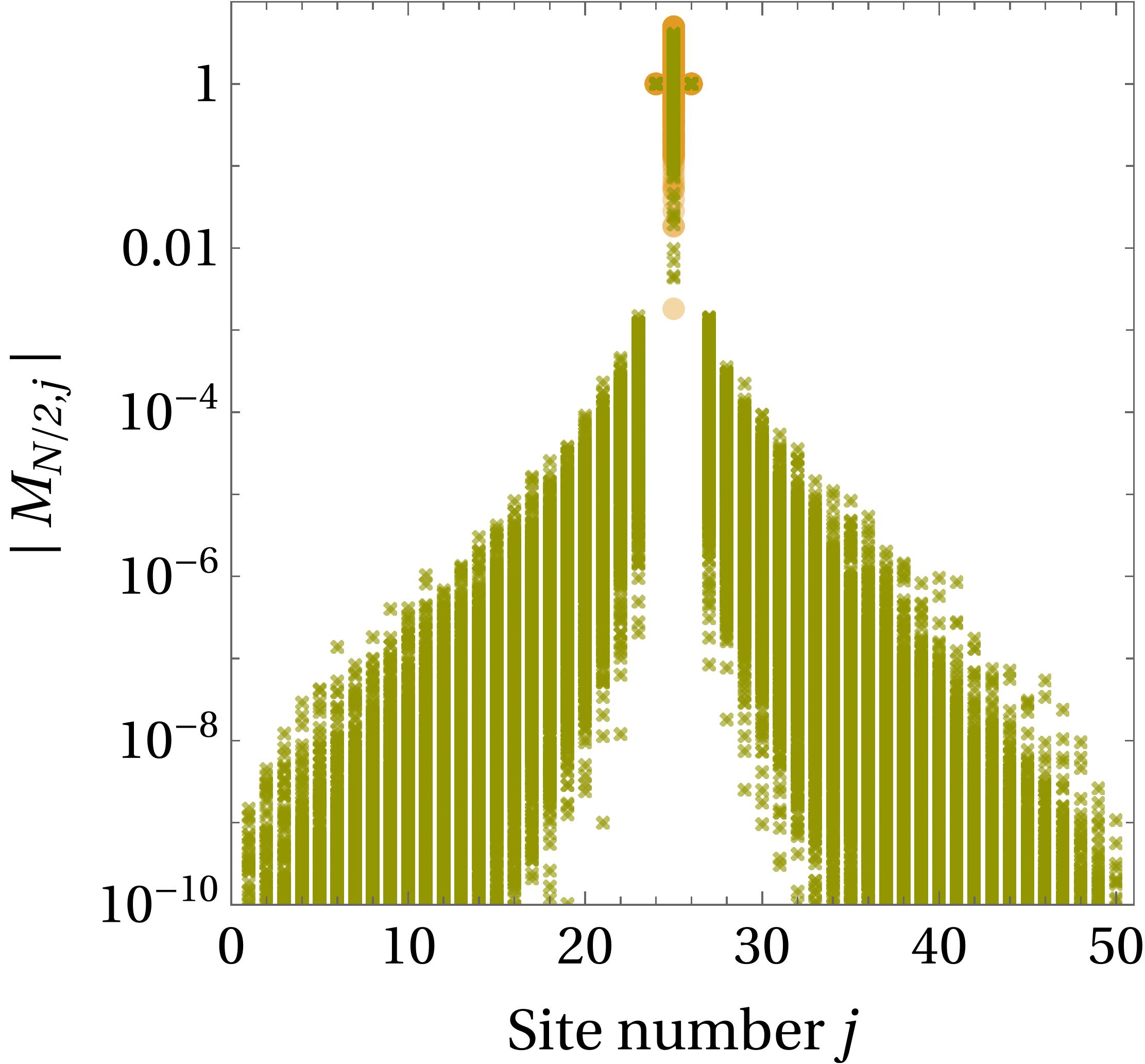}}\hspace*{0.5cm}
\subfigure{\includegraphics[width=0.43\textwidth]{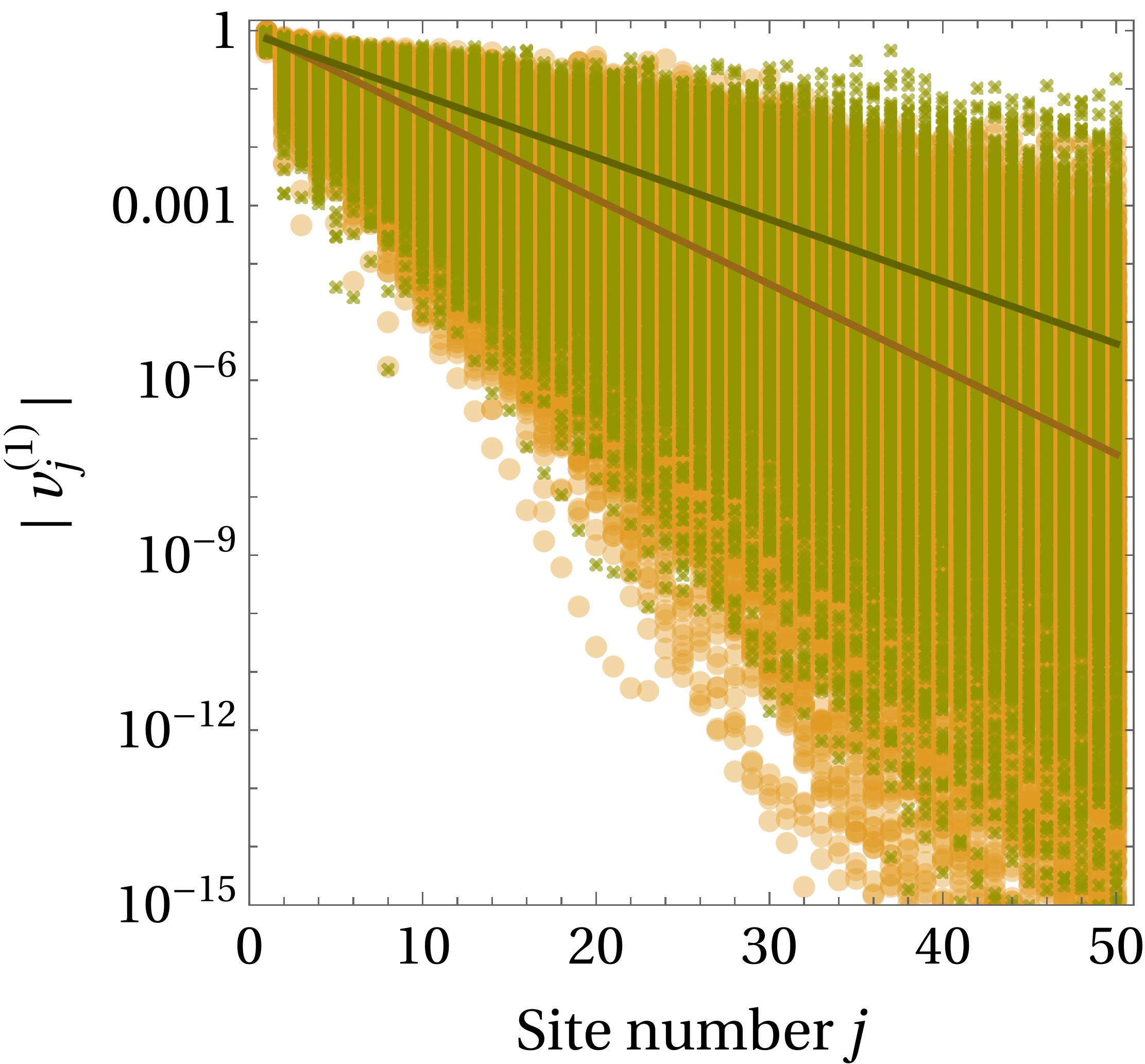}}
\caption{Same as Fig. \ref{fig4} but for the quantum corrections evaluated for the gauged version of quivers.}
\label{fig7}
\end{figure}

Substituting the charges of fermions in eq. (\ref{dM1}), the shift in the mass matrix in the present case is obtained as
\be \label{dM1_quiver}
\delta M_{ij} = \frac{g^2}{4 \pi^2}\,\sum_{a} {\cal R}_{ia} {\cal R}_{ja}\,\sum_{k}\, U^*_{ik}\,U^*_{jk}\,m_k\,B_0[M_a^2,m_k^2]\,.\ee
The gauge boson mixing matrix ${\cal R}$ can be determined by diagonalizing ${\cal M}_X^2$ and $M_a^2$ are the corresponding eigenvalues. For the form of ${\cal M}_X^2$ derived in eq. (\ref{M_GB}), one finds that ${\cal R}$ is different from the identity matrix and all its elements are of similar magnitude. This in turn implies the presence of the non-vanishing non-local terms in $\delta M_{ij} $. As usual, such effects are hierarchical and ordered as dictated by the localised leading order eigenvectors $v^{(k)}$. Simulating these effects numerically, we find the results as displayed in Fig. \ref{fig7}. 
Here, we have set $g v_\phi = 0.2 W$ while the rest of the details are identical to the one considered in the previous examples. It can be seen that the quantum corrections in this case increase the localisation length of the eigenvectors and at the same time induce hierarchical but non-vanishing non-local couplings in the Anderson chain. 

\section{Applications}
\label{sec:application}
As shown in the last two sections, the gauge interactions among the fermions in the chain can lead to the generation of non-nearest-neighbour couplings in the chain and modification in the localisation length of the eigenvectors through radiative corrections. The change in the localisation properties of the mass eigenstates has direct implications for the phenomenological applications considered earlier \cite{Craig:2017ppp}. Depending on the increase or decrease in the localisation length, the number of lattice sites needs to be decreased or increased to achieve the same exponential suppression between the fields attached at the opposite sides of the chain. The other novel set of phenomenological applications arises from the departure from the non-locality induced by the radiative effects. As outlined in section \ref{sec:massless}, when multiple fields are attached to a 1D chain in a specific manner, all except one of them stay massless due to the locality of interaction in the theory space. The radiative effects break this and hence can induce masses for these multiplets. Some applications of this variety are discussed below.

\subsection{Flavour hierarchies}
\label{subsec:flavour}
The most straightforward application of the underlying setup is in the flavour sector of the SM. For $N_f=3$, the arrangement sketched in Fig. \ref{fig3} leads to two massless and one massive generation at the leading order. This changes when the fermions are to have some additional interactions and the next-to-leading order corrections arising due to them are taken into account in the full mass matrix ${\cal M}$ given in eq. (\ref{M_fh}). As the simplest case, consider that such interactions are of the type that we considered in section \ref{subsec:universal}. Only the fermions in the 1D lattice chain are charged under the site-universal $U(1)$. The radiative corrections change $M \to M+\delta M$, while the other blocks of ${\cal M}$ remain unaltered in this case. The new $M$ possesses the same eigenvectors $v^{(k)}$ but with the eigenvalues $m_k$ changed to $m_k+\delta m_k$, where $\delta m_k$ is the shift which can be calculated in terms of the parameters of tree-level $M$ and the mass of the gauge boson and gauge coupling of the site-universal $U(1)$.

Integrating out the heavy fermions and following eqs. (\ref{M_eff},\ref{M_eff_ab}), the 1-loop corrected effective mass matrix for $N_f$ flavours, in this case, is obtained as
\beqa \label{M1_eff}
\left(m^{(1)}_{\rm eff} \right)_{\alpha \beta} &\simeq & - \mu_\alpha\, \mu_\beta^\prime\, \sum_{k=1}^N \frac{1}{m_k + \delta m_k}\, v_\alpha^{(k)}\, v_{N+1-\beta}^{(k)} \nonumber \\
& \approx & \left(m_{\rm eff} \right)_{\alpha \beta}   +  \mu_\alpha\, \mu_\beta^\prime\, \sum_{k=1}^N \frac{\delta m_k}{m_k^2}\, v_\alpha^{(k)}\, v_{N+1-\beta}^{(k)}\,.\eeqa
As before, the matrix $m_{\rm eff} $ is of rank one, and it gives rise to mass for the third generation. The second term induced by the one-loop corrections in the fermionic chain can induce masses for the other two generations. Therefore, the masses of these fermions are loop-suppressed in comparison to the mass of the third-generation fermion. The latter itself could be much smaller than the typical mass scale of chain fermions due to Anderson localisation, see eq. (\ref{Tr}).

The masses of the first two generations can also be hierarchical. The correction to the effective  matrix responsible for their masses can be written as,
\beqa \label{dM_eff}
\left(\delta m_{\rm eff} \right)_{\alpha \beta} &=&  \mu_\alpha\, \mu_\beta^\prime\, \sum_{k=1}^N \frac{\delta m_k}{m_k^2}\, v_\alpha^{(k)}\, v_{N+1-\beta}^{(k)} \nonumber \\
& \sim &  \mu_\alpha\, \mu_\beta^\prime\, \sum_{k=1}^N \frac{\delta m_k}{m_k^2}\, \left(v_{k_0}^{(k)}\right)^2\, \exp\left[-\frac{|\alpha - k_0| + |N-\beta+1-k_0|}{L_k}\right]\,.\eeqa
From the sum, the largest contribution typically comes from the term corresponding to the smallest value of the numerator under the exponent. For $\alpha < k_0 < N-\beta+1$, it is straightforward to see that such a value corresponds to $N+1-(\alpha + \beta)$. For $k_0 \leq \alpha \leq N-\beta+1$, one finds the numerator as $N+1-(\alpha + \beta) - 2 (k_0 - \alpha)$ and hence the minimum value it can take is again $N+1-(\alpha + \beta)$. The same can be shown for the remaining choice, $\alpha \leq N-\beta+1 \leq k_0$. Therefore, the typical strength of the $(\alpha \beta)^{\rm th}$ elements of $\delta m_{\rm eff}$ is given by 
\be \label{dm_eff_1}
\left(\delta m_{\rm eff} \right)_{\alpha \beta} \sim \exp\left[-\frac{N+1-(\alpha + \beta)}{L_k}\right]\,.\ee
The matrix $\delta m_{\rm eff}$ is hierarchical, and so are the masses of the first two generations. The intergenerational mass hierarchy among these fermions is controlled by the number of sites $N$ and localisation length $L_k$. A stronger hierarchy can be generated using sufficiently large $N$ for a given $L_k$.  

\begin{figure}[t]
\centering
\subfigure{\includegraphics[width=0.28\textwidth]{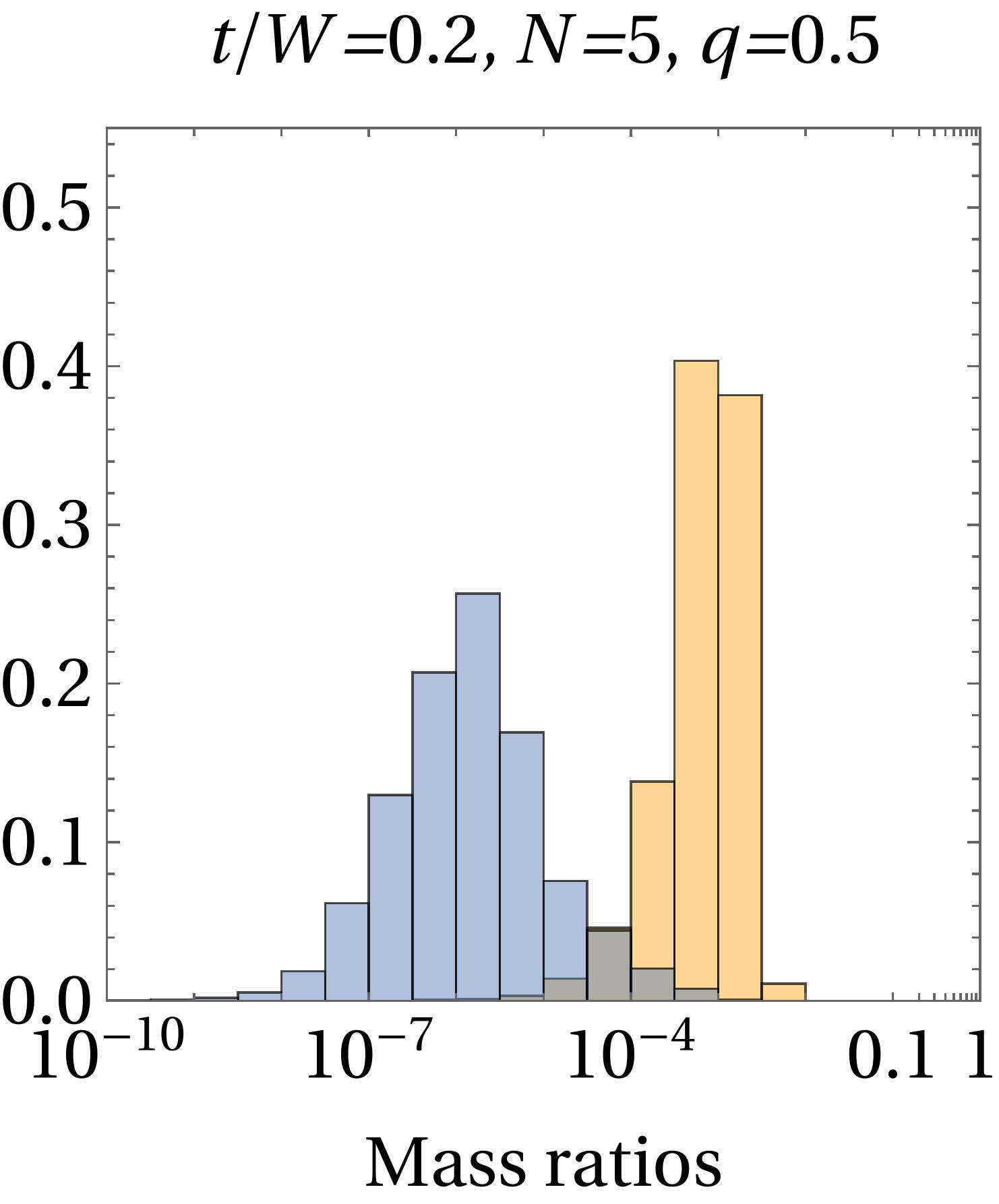}}\hspace*{0.5cm}
\subfigure{\includegraphics[width=0.28\textwidth]{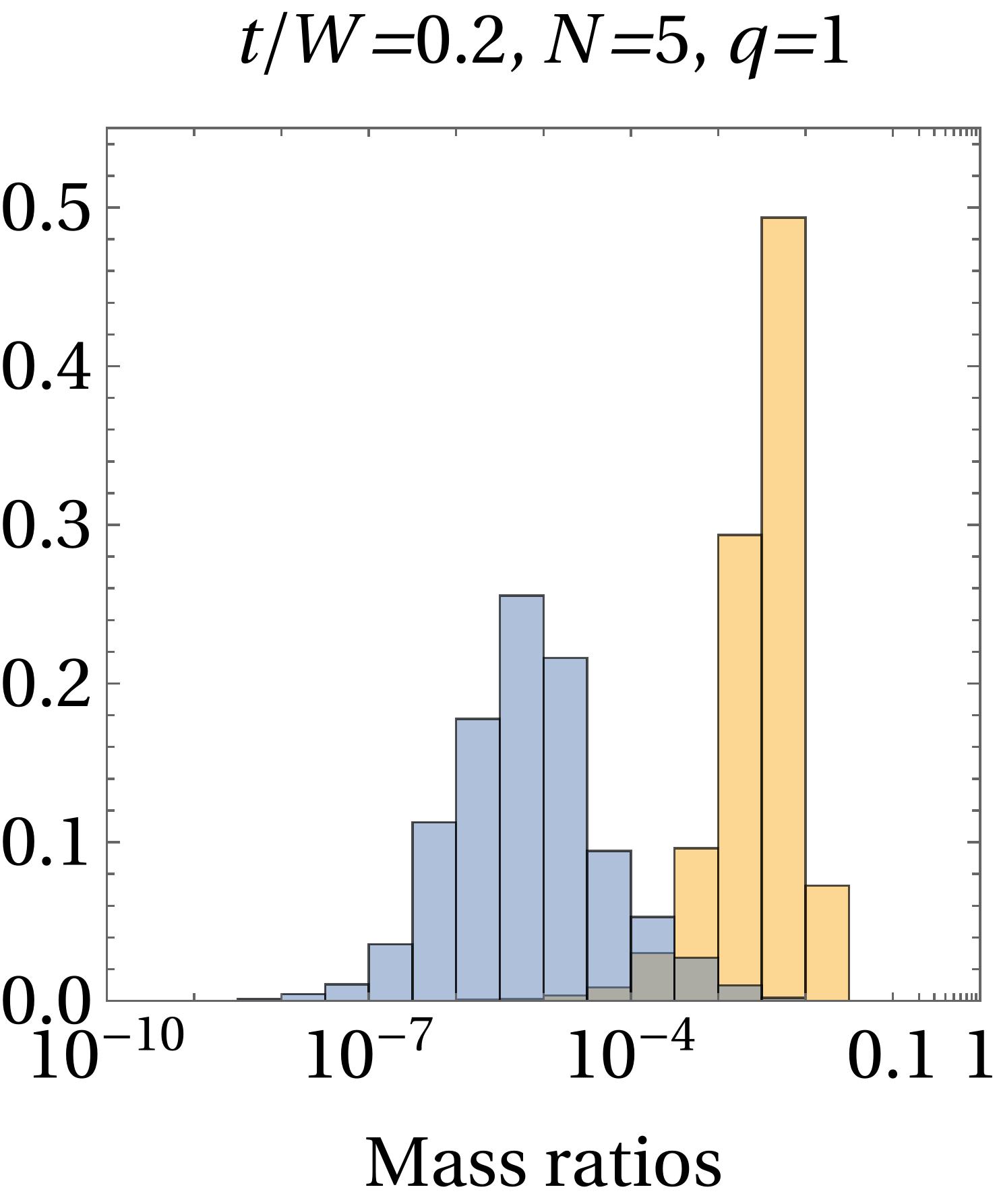}}\hspace*{0.5cm}
\subfigure{\includegraphics[width=0.28\textwidth]{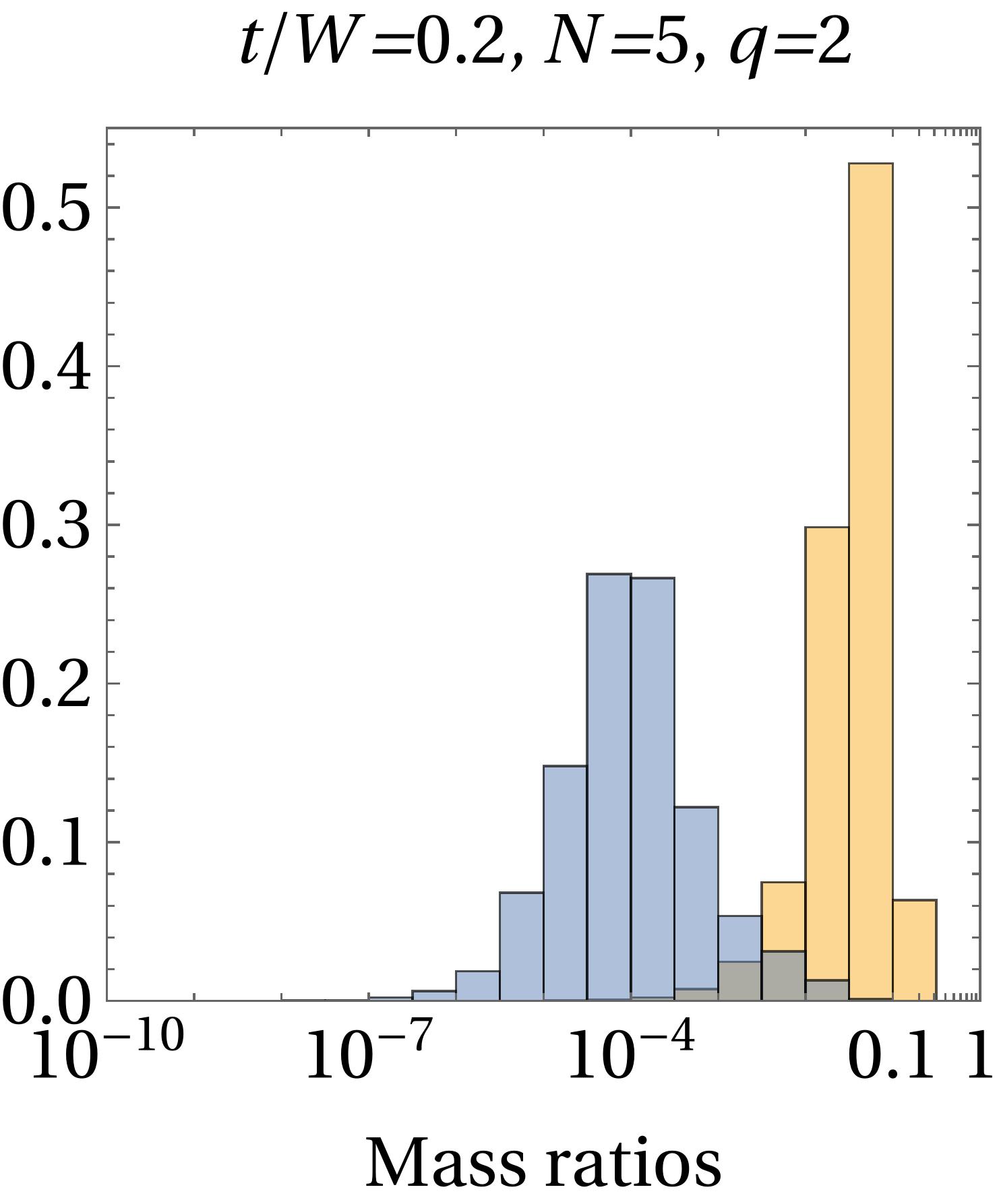}}\\
\subfigure{\includegraphics[width=0.28\textwidth]{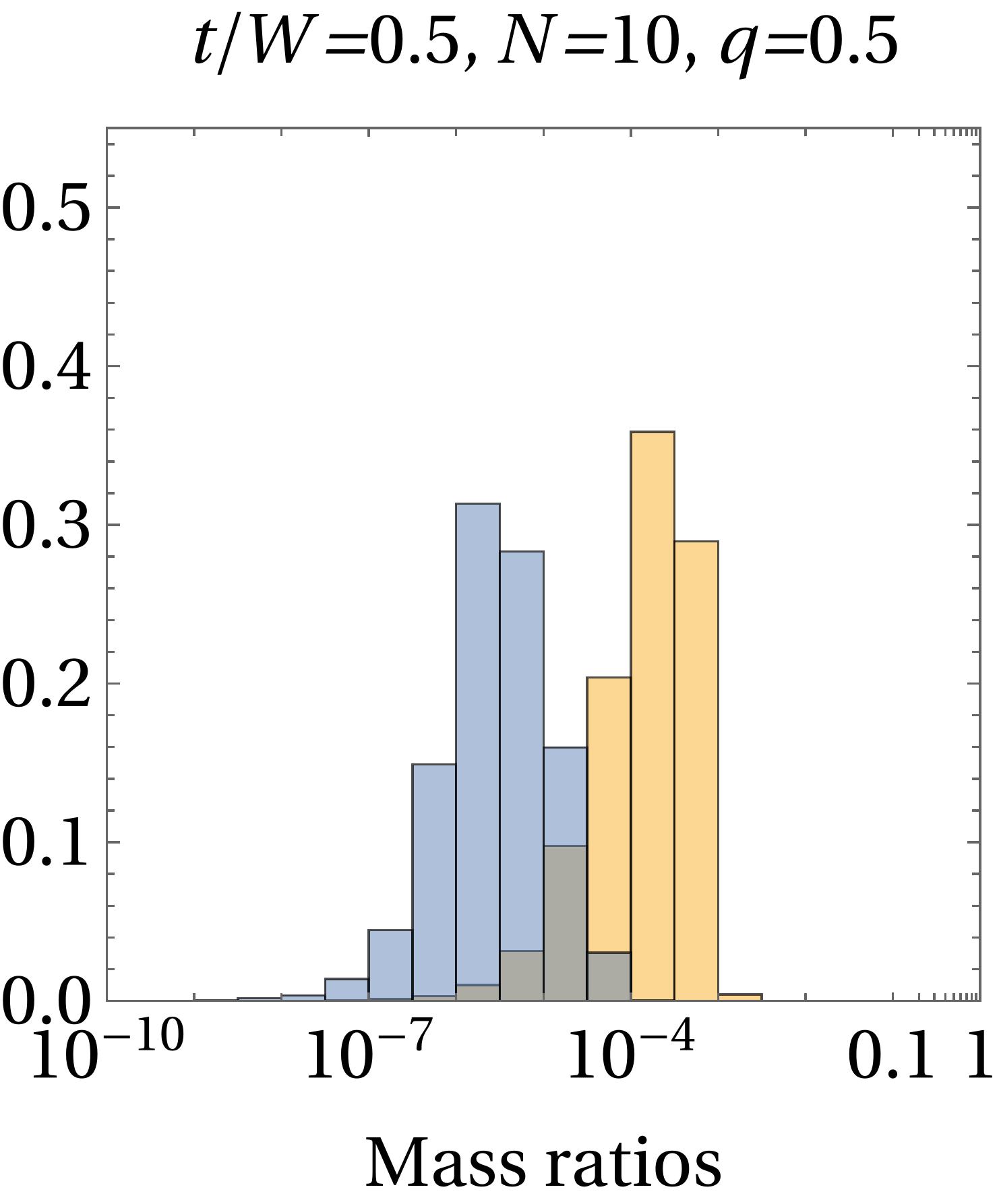}}\hspace*{0.5cm}
\subfigure{\includegraphics[width=0.28\textwidth]{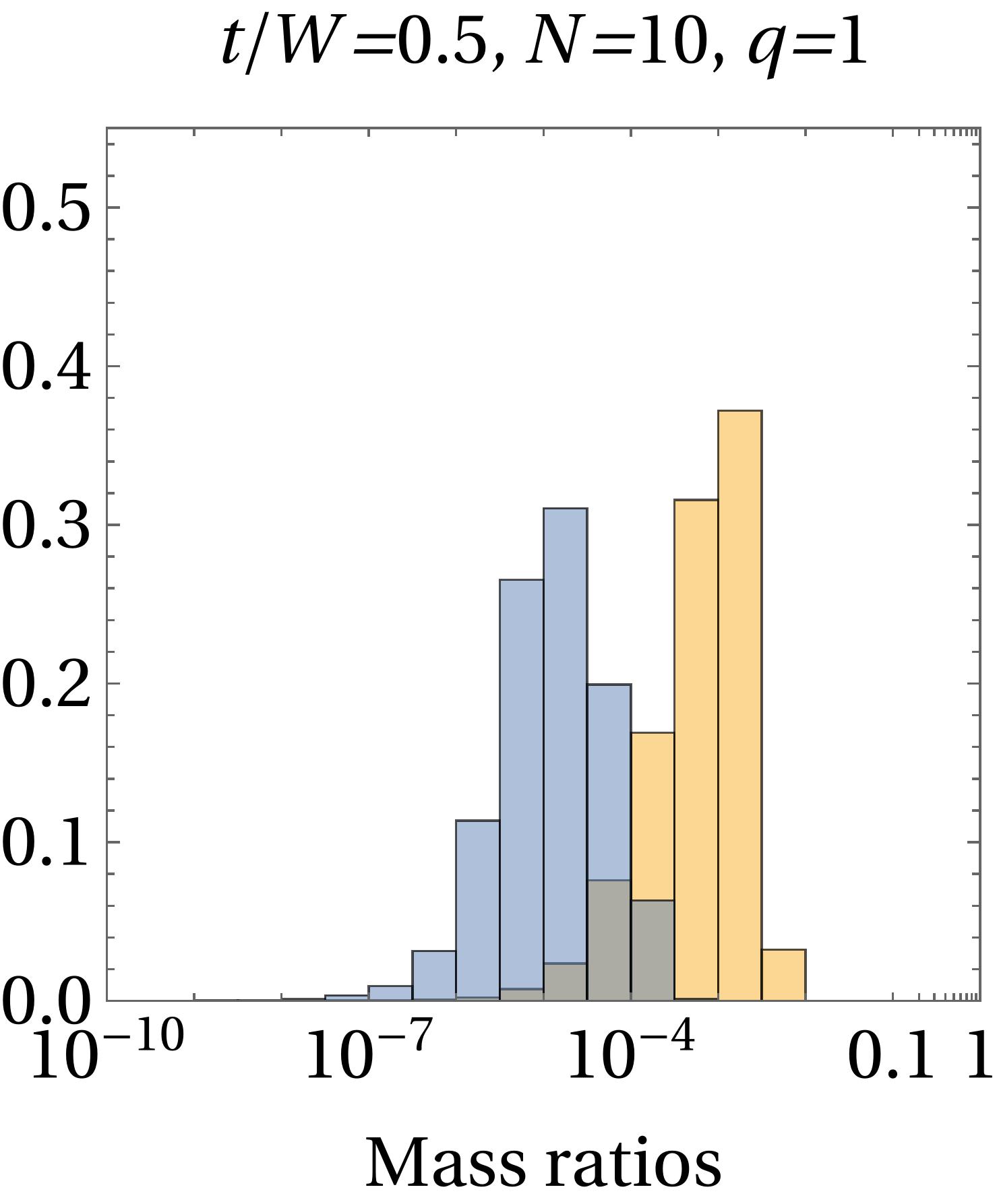}}\hspace*{0.5cm}
\subfigure{\includegraphics[width=0.28\textwidth]{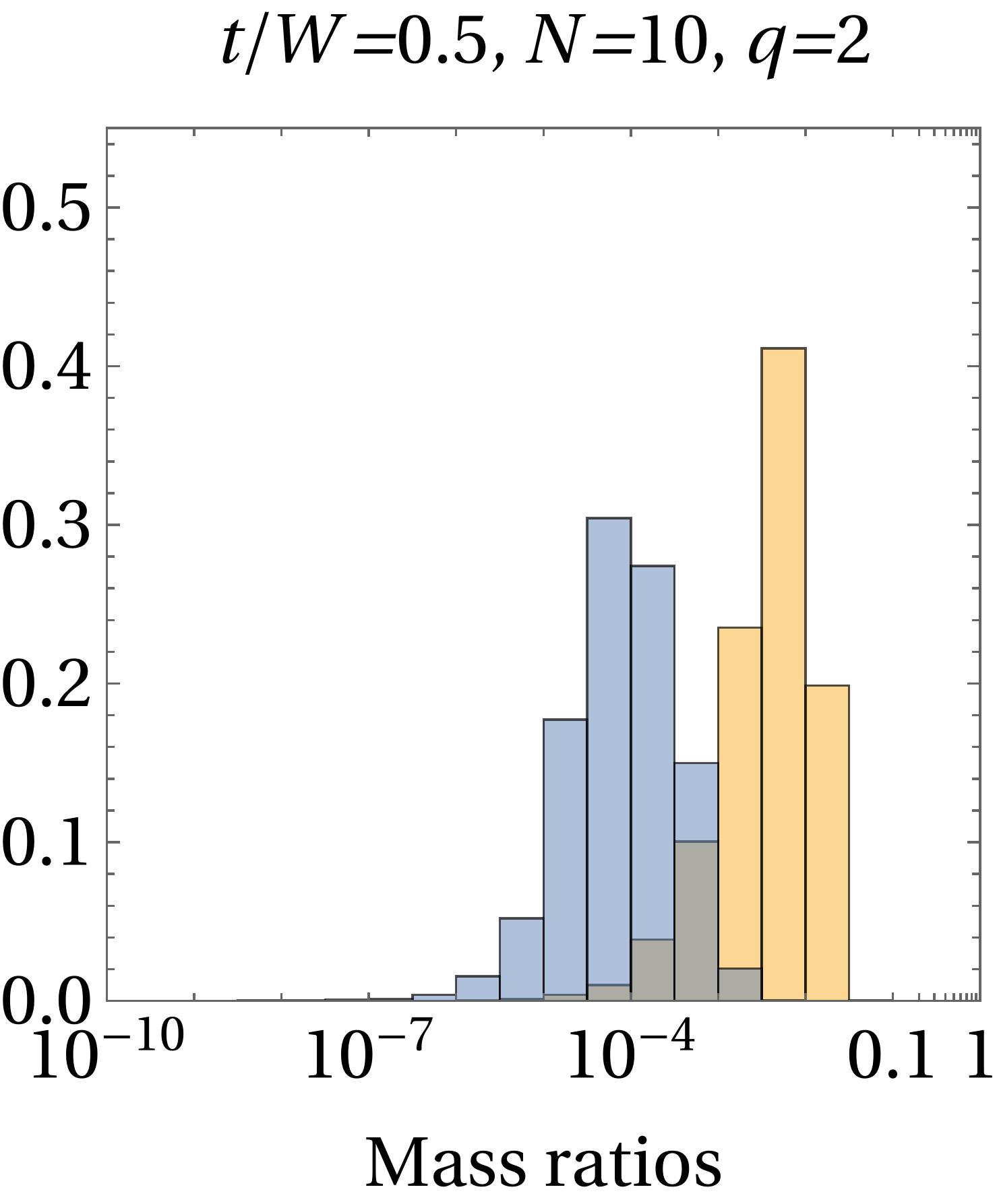}}\\
\caption{Some illustrative cases leading to intergenerational mass hierarchies of ${\cal O}(10^{-1})$ to ${\cal O}(10^{-6})$ for $N_f=3$. The histograms are of second-to-third and first-to-third generation masses and are plotted for $10^4$ samples.}
\label{fig8}
\end{figure}
For practical demonstration, we compute the intergenerational mass hierarchies for $N_f = 3$. We use $m_{\rm eff}$ given in eq. (\ref{M_eff}) with 1-loop corrected $M$ in the presence of site-universal $U(1)$ interactions. As before, the diagonal elements of $M$ are taken from a uniform random distribution over $[0, W]$ with $W=5$ TeV. Eq. (\ref{dM1_uni}) is used as a correction to the tree-level mass matrix, and we choose $M_X=0.2\, W$, $g=1$ as before, and consider three reference values for $q$. Choosing $\mu_\alpha=\mu_\alpha^\prime=M_Z$, we compute the ratios of second-to-third and first-to-third generation fermions. The results are displayed in Fig. \ref{fig8} for two sample values of $N$ and $t/W$ for which intergenerational hierarchies in the range ${\cal O}(10^{-1})$ to ${\cal O}(10^{-6})$ are obtained. It can be seen that the mass ratio between the second and third-generation fermions is primarily controlled by the strength of the loop correction. The intergenerational hierarchy between the first and second generations is independent of $gq$.  The two to three orders of magnitude difference between them can be obtained for the number of lattice sites as small as $N=5$.

The mechanism can be straightforwardly implemented into the SM to generate intergenerational mass hierarchies among the charged fermions. Multiple copies of vectorlike pairs with quantum numbers similar to those of $u_R$, $d_R$ and $e_R$ under the SM gauge symmetry, are to be introduced to form a 1D lattice chain. The mass parameters $\mu_\alpha$ can arise from the electroweak symmetry breaking when one or more copies of the SM Higgs acquire non-zero expectation values.  The parameters $\mu^\prime_\alpha$ can also be introduced as spurions of the new $U(1)$ under which only the chain fermions are charged. The top quark, bottom quark and tau lepton obtain their masses at tree level. The two orders of magnitude hierarchy between the top and bottom quarks can be introduced by making appropriate choices of $N$ for them. The first and second-generation fermions receive their masses once the quantum corrections in the fermionic chain are enabled. Explicit models and their phenomenological implications such as flavour mixing, flavour violation induced by mixing with vectorlike fermions, a lower limit on the scale $W$, etc. demand detailed investigations, and they would be separately studied. 

It is noteworthy that the above mechanism is qualitatively different from the one proposed in \cite{Craig:2017ppp} in the context of simple Anderson localisation. The latter uses the exponential profile of a localised scalar and partial compositeness to give rise to the flavour hierarchies. The role played by the strong dynamics in the composite sector in the usual model of this type \cite{Kaplan:1991dc} is replaced by Anderson localisation of scalars, and an efficient arrangement requires a chain consisting of $N \sim 9$ scalars. The mechanism we propose here is conceptually distinct and also exhibits unique features from a phenomenological perspective.

\subsection{Neutrino mass}
\begin{figure}[t!]
\centering
\subfigure{\includegraphics[width=0.85\textwidth]{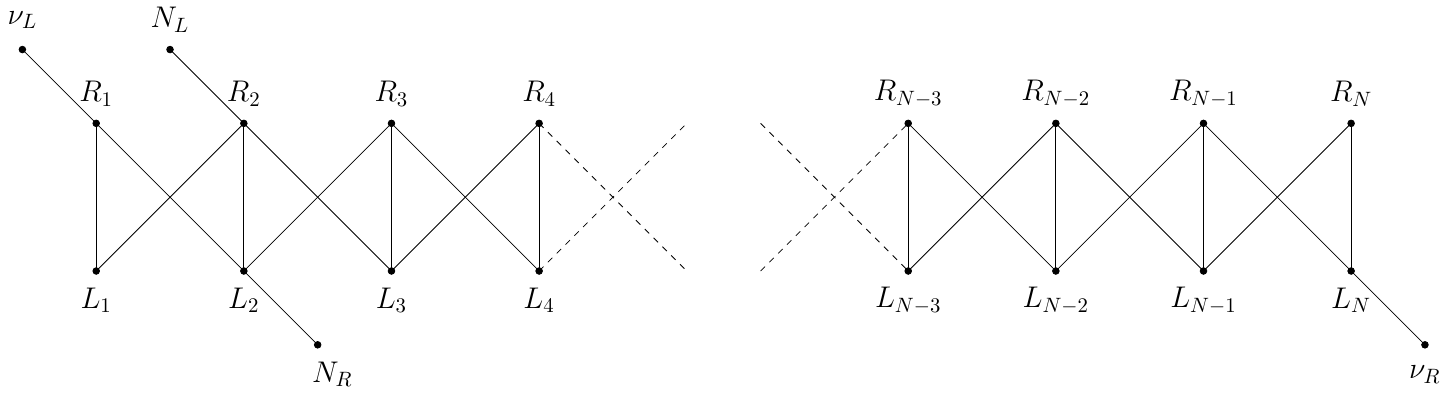}}
\caption{An example of neutrino fields attachment to the Anderson chain leading to tiny neutrino masses.}
\label{fig9}
\end{figure}

Small neutrino mass can also be obtained as a consequence of radiatively induced non-locality in the Anderson chain. Consider a one-generation case in which the SM lepton doublet and the right handed neutrino $\nu_R$ are attached to the fermionic chain at the opposite ends. We introduce a pair of SM singlets, $N_{L,R}$, which are attached at the same site in the chain, see Fig. \ref{fig9}. 
The corresponding mass terms can be written as 
\be \label{L_nu}
-{\cal L} \supset \mu_\nu\,\overline{\nu}_L R_1+\mu_N\,\overline{N}_L R_2+\mu_N^\prime\,\overline{L}_2 N_R + \mu_\nu^\prime\,\overline{L}_N \nu_R+ \sum_{i,j=1}^N\,M_{ij}\,\overline{L}_i R_j\,+\,{\rm h.c.}\,.\ee
Integrating out the fermions in the chain, the effective Dirac matrix $M_D$ for the remaining ones, in the basis $n_{L,R}=\left(\nu_{L,R},N_{L,R}\right)^T$ can be obtained as 
\be \label{MD}
M_D = -\left(\ba{cc} \mu_\nu \mu_\nu^\prime (M^{-1})_{1N} & \mu_\nu \mu_N^\prime (M^{-1})_{12} \\ \mu_N \mu_\nu^\prime (M^{-1})_{2N} & \mu_N \mu_N^\prime (M^{-1})_{22} \ea\right)\,,\ee
where, as usual, the elements of $M^{-1}$ can be expressed as
\be \label{Min_neut}
\left(M^{-1}\right)_{ij} = \sum_{k=1}^N\,\frac{1}{m_k}\,v^{(k)}_i\,v^{(k)}_j\,,\ee
with the help of eq. (\ref{UMU}).

If one uses the tree-level expression of $M$ then $M_D$ can be shown to possess one massless state following the result of section \ref{sec:massless}. Explicitly, using the relation for $v^{(k)}_2$ given in eq. (\ref{vk_iter}) in (\ref{Min_neut}), we can write
\beqa \label{Minv_rel}
\left(M^{-1}\right)_{2N} &=&  \sum_{k=1}^N\,\frac{1}{m_k}\,v^{(k)}_2\,v^{(k)}_N = -\frac{\epsilon_1}{t} \left(M^{-1}\right)_{1N}\,, \nonumber \\
\left(M^{-1}\right)_{22} &=&  \sum_{k=1}^N\,\frac{1}{m_k}\,v^{(k)}_2\,v^{(k)}_2 = -\frac{\epsilon_1}{t} \left(M^{-1}\right)_{12}\,.
\eeqa 
Substituting the above in $M_D$, one finds that the first and second rows are related by simple scaling. Hence, one of the eigenvalues is zero while the other is of ${\cal O}(\mu_N \mu_N^\prime/W)$.

The massless state obtains suppressed masses when the quantum effects in the fermionic chain are considered. Assuming again the site-universal $U(1)$, the shift in $M_D$ induced by 1-loop corrections in $M$ are obtained as
\be \label{dMD}
\delta M_D = \left(\ba{cc} \mu_\nu \mu_\nu^\prime \sum_k\frac{\delta m_k}{m_k^2} v^{(k)}_1 v^{(k)}_N & \mu_\nu \mu_N^\prime \sum_k\frac{\delta m_k}{m_k^2} v^{(k)}_1 v^{(k)}_2 \\ \mu_N \mu_\nu^\prime \sum_k\frac{\delta m_k}{m_k^2} v^{(k)}_2 v^{(k)}_N & \mu_N \mu_N^\prime \sum_k\frac{\delta m_k}{m_k^2} v^{(k)}_2 v^{(k)}_2\ea\right)\,.\ee
The induced non-zero mass can be identified with that of the light neutrino, and it can be approximated as
\be \label{mnu_D}
m_\nu \approx \mu_\nu \mu^\prime_\nu \sum_k \frac{\delta m_k}{m_k^2} v^{(k)}_1 v^{(k)}_N \sim \mu_\nu \mu^\prime_\nu \sum_k \frac{\delta m_k}{m_k^2}\, e^{-\frac{N-1}{L_k}}\,.\ee
The light neutrino mass is suppressed by both the loop effects and Anderson localisation.

\begin{figure}[t]
\centering
\subfigure{\includegraphics[width=0.28\textwidth]{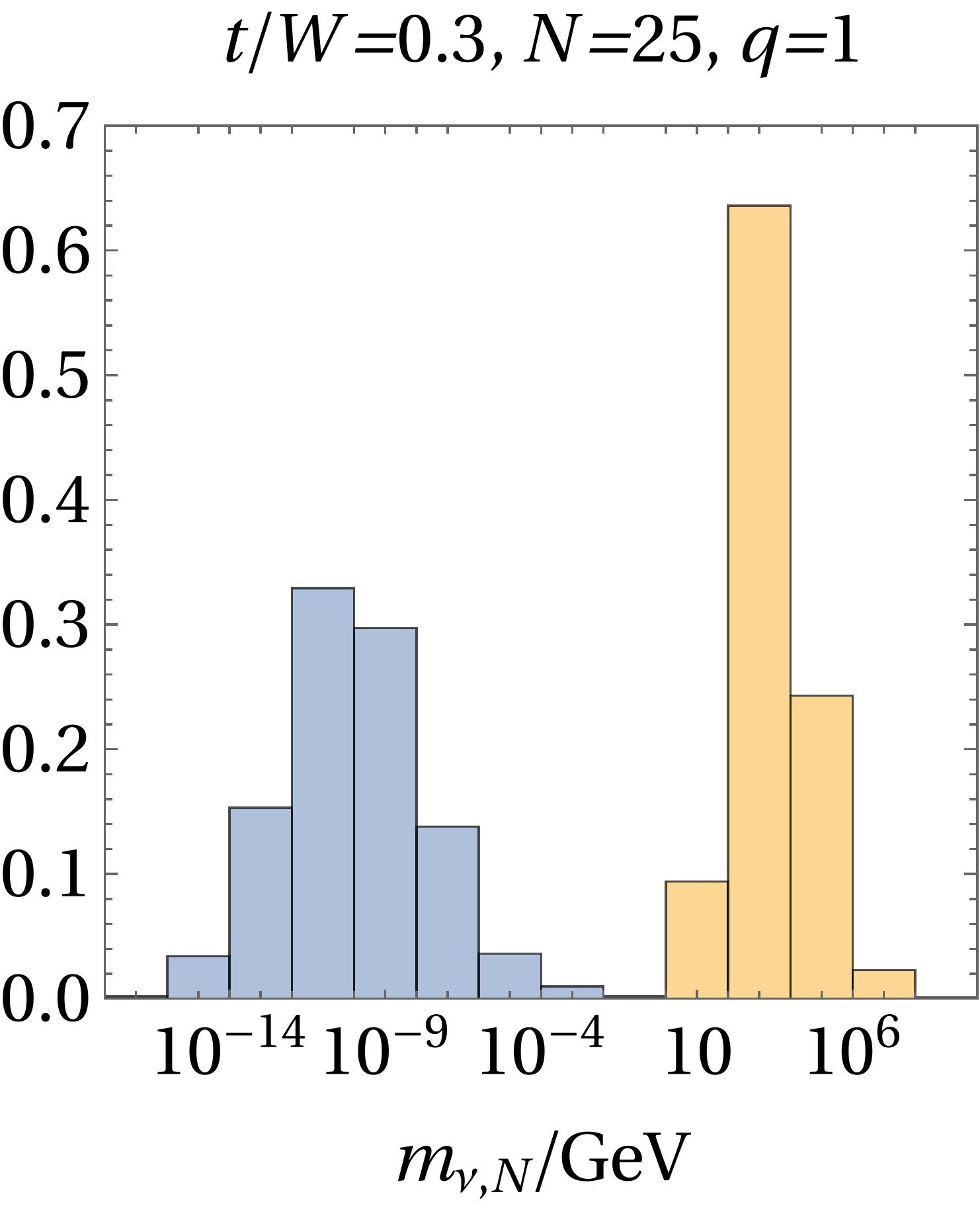}}\hspace*{0.5cm}
\subfigure{\includegraphics[width=0.28\textwidth]{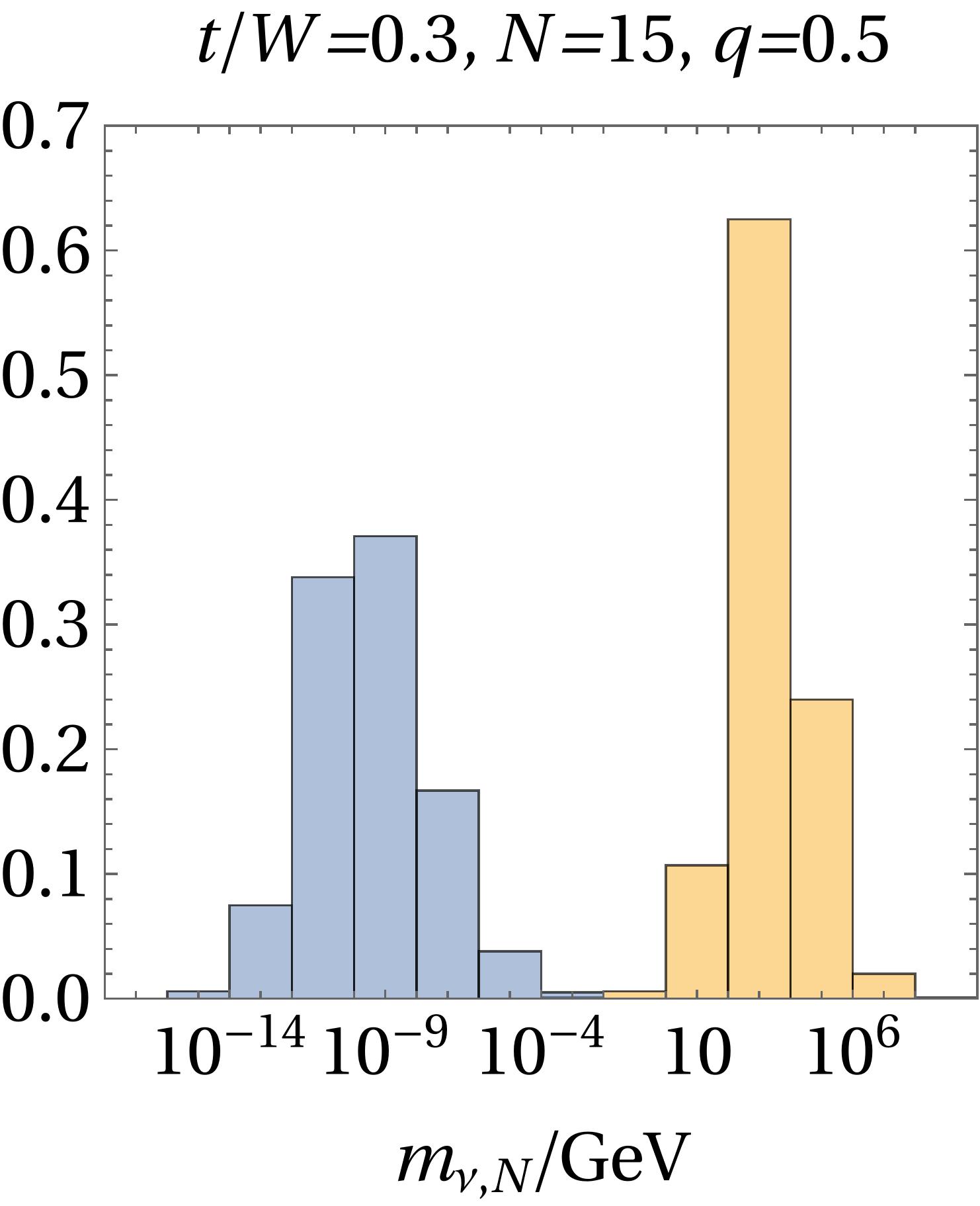}}\hspace*{0.5cm}
\subfigure{\includegraphics[width=0.28\textwidth]{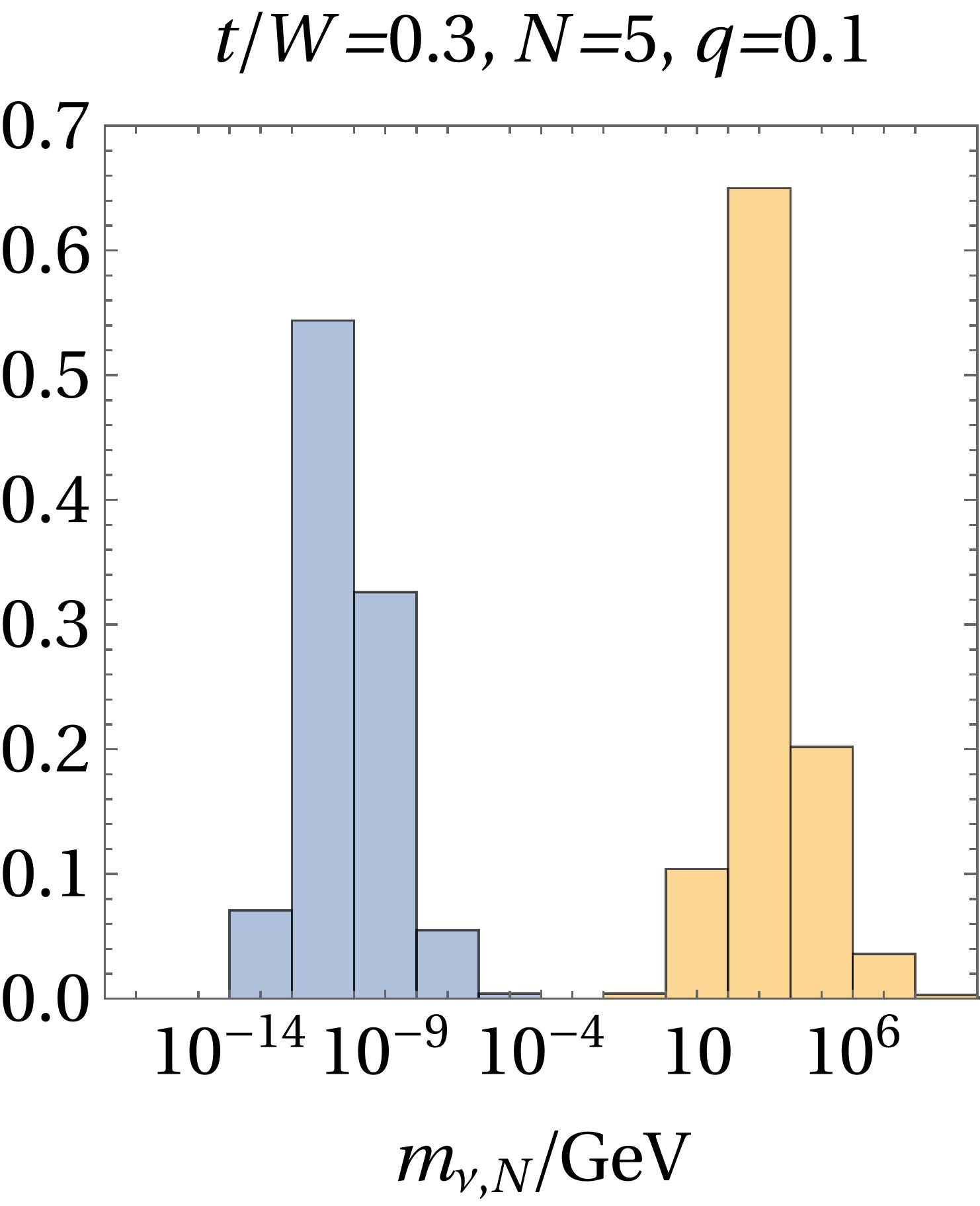}}\\
\caption{The masses of $\nu$ and $N$ states computed from $M_\nu$ given in eq. (\ref{Mnu}) for different values of $N$, $q$ and $10^3$ samples of $M$. The quantum corrections in the latter are considered in the presence of site-universal abelian gauge interaction.}
\label{fig10}
\end{figure}
To compare quantitatively with the neutrino mass generation mechanism presented in the context of the usual Anderson mechanism in \cite{Craig:2017ppp}, we also introduce Majorana masses for the singlets $N_R$ and $\nu_R$. For simplicity, they are considered to have the same mass of the order of $W$ and no mixing, i.e. 
\be \label{Majorana}
-{\cal L} \supset \frac{1}{2}\,W\,\left(\overline{\nu_R^c}\, \nu_R +  \overline{N_R^c}\, N_R\right)\,.\ee
This and the Dirac mass matrix $M_D^{(1)} =M_D + \delta M_D$ leads to the effective mass matrix for $\nu$ and $N$ states as
\be \label{Mnu}
M_\nu = - \frac{1}{W}\,M_D^{(1)}\,M_D^{(1) T}\,.\ee
The Majorana mass of the lightest state is now given by
\be \label{mnu_light}
m_\nu \sim \frac{1}{W}\left(\mu_\nu \mu^\prime_\nu \sum_k \frac{\delta m_k}{m_k^2}\, e^{-\frac{N-1}{L_k}}\right)^2\,.\ee

The additional suppression in $m_\nu$ arranged by the loop effects allows one to consider relatively small $N$ in comparison to the usual Anderson mechanism. For the demonstration, we consider $\mu_\nu = \mu_\nu^\prime = M_Z$, $\mu_N=\mu_N^\prime = W/2$, $t/W=0.3$,  $W=5$ TeV and numerically obtain the masses of $\nu$ and $N$ states for  some example values of $N$ and $q$. The results are shown in Fig. \ref{fig10}. Consequently, the sub-eV scale neutrino mass can be produced even for $N=5$ if $q=0.1$. This is much smaller in comparison to $N \approx 30$ needed in the case of ordinary Anderson localisation \cite{Craig:2017ppp}.

\subsection{Higgs mixing and $\mu$ problem}
Another example in which massless modes emerging from the locality can be utilised for phenomenological application is supersymmetric theories involving multiple copies of the electroweak doublet chiral superfields $\hat{H}^{u,d}$. In the frameworks based on grand unification, more than one copy of $\hat{H}^{u,d}$ arise from different representations of the unified gauge group with the typical mass scale of the order of the unification scale $\sim M_{\rm GUT}$. These multiplets are required to mix to reproduce viable quark and lepton masses and mixing angles \cite{Babu:1992ia,Aulakh:2003kg,Joshipura:2011nn}. Moreover, at least a pair of their combinations is required to remain light so that it can induce the breaking of electroweak symmetry at a scale much below $M_{\rm GUT}$. This is similar to the well-known $\mu$ problem in supersymmetric versions of the SM with additional complicity of the presence of multiple copies of $\hat{H}^{u,d}$.

It is straightforward to solve this problem in the present class of theories by considering the supersymmetric version of Fig. \ref{fig9}. For simplicity, we consider only two copies, $\hat{H}^u_{1,2}$ and $\hat{H}^d_{1,2}$, to demonstrate this mechanism, which can be straightforwardly generalised to multiple copies too. The fermions in the chain are replaced by the chiral superfields $\hat{h}^u_i$ and $\hat{h}^d_i$ with the  SM quantum numbers identical to those of $\hat{H}^u$ and $\hat{H}^d$, respectively. Attaching $\hat{H}^{u,d}_1$ at the opposite ends of the chain and $\hat{H}^{u,d}_2$ at the same site, the superpotential can be arranged as
\be \label{W}
{\cal W} \supset \mu_1\,\hat{H}^{u}_1 \hat{h}^d_1+\mu_2\,\hat{H}^{u}_2 \hat{h}^d_2+\mu_2^\prime\,\hat{h}^u_2 \hat{H}^d_2 + \mu_1^\prime\,\hat{h}^u_N \hat{H}^d_1+ \sum_{i,j=1}^N\,M_{ij}\,\hat{h}^u_i \hat{h}^d_j\,.\ee
Again, integrating out the chiral superfields in the chain, one obtains the effective, $\mu_{\alpha \beta} H^{u}_\alpha H^d_\beta$, with $\mu_{\alpha \beta} $ are the elements of $2 \times 2$ matrix,
\be \label{MU}
{\cal \mu} = \left(\ba{cc} -\mu_1 \mu_1^\prime (M^{-1})_{1N} & -\mu_1 \mu_2^\prime (M^{-1})_{12} \\ \frac{\epsilon_1}{t}\mu_2 \mu_1^\prime (M^{-1})_{1N} & \frac{\epsilon_1}{t}\mu_2 \mu_2^\prime (M^{-1})_{12} \ea\right)\,.\ee
Here, we have already used the relations, eq. (\ref{Minv_rel}), applicable for the tree-level $M$.

The consequence of the above matrix is that it has a vanishing eigenvalue, leading to a pair of the combination of $\hat{H}^{u,d}_{1,2}$ with vanishing $\mu$ term. From the diagonalisation of the above matrix, the massless combinations are obtained as
\beqa \label{Hud}
\hat{H}^{u,d} &=& \cos \theta_{u,d}\,\hat{H}_{1}^{u,d} + \sin \theta_{u,d}\,\hat{H}_{2}^{u,d}\,.\eeqa
with
\beqa \label{th_ud}
\tan 2\theta_u &\simeq & \frac{2 \epsilon_1 \mu_2}{t \mu_1} \left(\frac{\epsilon_1^2 \mu_2^2}{t^2 \mu_1^2} -1\right)^{-1}\,, \nonumber \\
\tan 2\theta_d &\simeq &  -\frac{2 \mu_1^\prime (M^{-1})_{1N}}{\mu_2^\prime (M^{-1})_{12}}\,.\eeqa
For generic values of underlying parameters, it can be seen that $\hat{H}^u$ is a well-mixed state while $\hat{H}_d$ has a dominant component of $\hat{H}^{d}_1$ for large $N$. This is because the $\hat{H}^u_{1,2}$ are closely spaced in comparison to $\hat{H}^d_{1,2}$ on the chain in the theory space. The mixing can be manipulated not only by $N$ but also by changing the sites to which $\hat{H}^{u,d}_2$ are to be attached. In analogy with the previous example, it can be also seen that the other combination of the chiral superfields obtains the mass of ${\cal O}(\mu_2 \mu_2^\prime/W)$ which is of the order of $M_{\rm GUT}$ if all the parameters in eq. (\ref{W}) are of a similar order.

The above setting delivers a pair $\hat{H}^{u,d}$ with accidentally vanishing $\mu$-term. The non-zero value of the latter may arise through the non-local effects. However, such effects in this case need not arise from the additional interactions between the chiral superfields in the chain. The nonrenormalisation theorem \cite{Grisaru:1979wc,Seiberg:1993vc} ensures that the corrections to $M$ vanish in the exact supersymmetric limit. The non-local effects are therefore protected by supersymmetry and can arise only when the latter is broken. For example, the non-zero $\mu$ can arise once supersymmetry is broken by gravity mediation as a generic feature of these types of models, as shown recently in \cite{Dvali:2020uvd}. In the usual gravity-mediated scenarios, the supersymmetry breaking scale is set by the gravitino mass $m_{3/2}$ and the soft terms are generically expected to be ${\cal O}(m_{3/2})$. When supersymmetry is broken in this way, it universally induces the shifts of heavy fields of the order of $m_{3/2}$ which in turn can generate the $\mu$ term of the same size.

The setup naturally provides a massless pair of Higgs doublet superfields at a scale $W \gg m_{3/2}$ as desired. In general, the light states are an admixture of multiple states with the same quantum number present in the setup. Although we outlined a simple case of two multiplets, the generalisation to more $\hat{H}^{u,d}_i$ is straightforward. One pair needs to be attached to the opposite ends of the chain, while each of the remaining ones can be at the same sites. The exact amount of admixture in the massless states depends on the placement of Higgs doublet superfields on the chain. In this way, the number of lattice sites $N$ controls the mixing.

\section{Conclusion}
\label{sec:concl}
Theory space Anderson localisation is a useful mechanism to generate large hierarchies among the effective couplings from generic and disordered fundamental parameters. The only order it requires is the nearest-neighbour coupling structure among the quantum fields. Such locality in theory space resembles the one-dimensional lattice structure of Anderson's tight-binding model. This work shows that the mechanism in its most original form can be utilised to obtain multiple massless modes of the fields if the latter are attached to the 1D lattice in a particular manner but still with generic couplings. The vanishing mass for these modes is entirely due to the strict locality of interactions in the 1D chain. 

It is shown that even if the locality is introduced in the tree-level QFT Lagrangian, the higher-order corrections can break it depending on the nature of additional interactions possessed by the fields forming the chain. Such quantum effects are explicitly shown using a few examples of interactions arising from the spontaneously broken abelian gauge symmetries. Typically, the non-local terms are suppressed by the localisation of the eigenvectors itself along with the loop factor.   Specific cases of gauge interactions are discussed in which such non-local effects keep the Anderson localisation intact or even lead to improvement. Most importantly, the deviation from the locality gives rise to tiny masses for the otherwise massless modes.

The proposed mechanism finds natural applications in the setups in which one or multiple copies of states are expected to be much lighter in comparison to the overall scale of the theory. The suppression is not only arranged by the Anderson localisation, but it is also assisted by the small non-local effects in the theory space. One of the most interesting applications mentioned here is in the flavour sector of the standard model. The arrangement results in a massive third generation, while the small masses of the first and second generations at low energies arise entirely from quantum corrections in the ultraviolet setup. Conceptually, this provides a novel mechanism for the radiative generation of flavour hierarchies, in contrast to the ideas proposed in \cite{Georgi:1972hy,Mohapatra:1974wk,Barr:1978rv,Wilczek:1978xi,Yanagida:1979gs,Barbieri:1980tz,Balakrishna:1987qd,Mohanta:2022seo}, where quantum corrections are considered in the infrared. An explicit model for this framework and its phenomenological implications will be explored separately, following the approach of \cite{Mohanta:2022seo}. The present framework is based on fermions at lattice sites, but the same can be investigated for the scalars and gauge bosons and the potential applications can be worked out.

\section*{Acknowledgements}
I thank Anjan S. Joshipura, Udit Khanna, Namit Mahajan and Navinder Singh for insightful discussions, and Gurucharan Mohanta for going through the draft and offering useful comments. This work is supported by the Department of Space (DOS), Government of India. I also acknowledge partial support under the MATRICS project (MTR/2021/000049) from the Science \& Engineering Research Board (SERB), Department of Science and Technology (DST), Government of India.

\appendix
\section{Proof of ${\rm Rank} (m_{\rm eff})=1$}
\label{app:rank}
The $N_f \times N_f$ effective mass matrix $m_{\rm eff}$ derived in eq. (\ref{M_eff_ab}) can be rewritten as
\be \label{Meff_def_app}
\left(m_{\rm eff} \right)_{\alpha \beta} \simeq - \mu_\alpha\, \mu_\beta^\prime\, \lambda_{\alpha \beta}\,,\ee
with
\be \label{lambda_app}
\lambda_{\alpha \beta} = \sum_{k=1}^N \frac{1}{m_k}\, v_\alpha^{(k)}\, v_{N+1-\beta}^{(k)}\,.\ee
The $\lambda_{\alpha \beta}$ are entirely determined from the eigenvalues and eigenvectors of the matrix $M$. For the tree-level $M$ which contains only on-site and nearest-neighbour couplings, the recursive relation between the elements of eigenvectors, in terms of the elements of $M$ and the eigenvalues, is obtained as eq. (\ref{vk_recursive}). We use this to prove that the rank of $m_{\rm eff}$ is one for $N \geq 2N_f-1$ through the principle of mathematical induction.

Consider first $N_f=2$ which is the smallest non-trivial value of $N_f$. Using the recursion relation, one finds,
\beqa \label{vk_iter}
v^{(k)}_1 &\neq & 0\,, \nonumber \\
v^{(k)}_2 &=& -\left(\frac{\epsilon_1 - m_k}{t} \right)\,v^{(k)}_1\,, \eeqa
as well as
\beqa \label{vkN_iter}
v^{(k)}_N &\neq & 0\,, \nonumber \\
v^{(k)}_{N-1} &=& -\left(\frac{\epsilon_N - m_k}{t} \right)\,v^{(k)}_N\,. \eeqa
Using the above, we can express one of the rows or columns in terms of the other. For example,
\beqa \label{l12_l22}
\lambda_{21} &=& \sum_{k} \frac{1}{m_k}\, v^{(k)}_2\, v^{(k)}_{N} = -\frac{\epsilon_1}{t}\,\lambda_{11}\,, \nonumber \\
\lambda_{22} &=& \sum_{k} \frac{1}{m_k}\, v^{(k)}_2\, v^{(k)}_{N-1} = -\frac{\epsilon_1}{t}\,\lambda_{12}\,.\eeqa
Here, we have used the orthogonality of the eigenvectors and $N>2$. The scaling between each of the elements of the first and second rows of the matrix $\lambda$ is universal. This, through eq. (\ref{Meff_def_app}), implies that there is only one independent row in $m_{\rm eff}$. Hence,  ${\rm Rank} (m_{\rm eff})=1$.

Let's assume that $m_{\rm eff}$, or equivalently the matrix $\lambda$, has only one independent row or column for $N_f=n$. Therefore, the elements of the second-last row (column) of $\lambda$ can be expressed in terms of those of the last row (column) as
\be \label{nn_rel}
\lambda_{n-1,\alpha} = \kappa\, \lambda_{n,\alpha}\,,~~~\lambda_{\alpha,n-1} = \kappa^\prime\, \lambda_{\alpha,n}\,, \ee
for $\alpha = 1,...,n$. Note that $\kappa=\kappa^\prime$ if the matrix $\lambda$ is symmetric.

Next, we need to prove that for $N_f=n+1$ the $(n+1)^{\rm th}$ row or column of $\lambda$ is not independent, and it is essentially the scaled version of $n^{\rm th}$ row or column, respectively.  Let's begin with the first $n$ elements of the last row. For $\alpha=1,...,n$, we can express
\beqa \label{row_rel}
\lambda_{n+1,\alpha} &=& \sum_k\,\frac{1}{m_k}\,v_{n+1}^{(k)}\, v_{N+1-\alpha}^{(k)}\, \nonumber \\ 
&=& \sum_k\,\frac{1}{m_k}\,\left[-v_{n-1}^{(k)}-\left(\frac{\epsilon_n-m_k}{t} \right) v_n^{(k)}\right]\,v_{N+1-\alpha}^{(k)}\, \nonumber \\  
&=& -\lambda_{n-1,\alpha} - \frac{\epsilon_n}{t}\,\lambda_{n,\alpha}\, \nonumber \\  
&=&  -\left(\kappa + \frac{\epsilon_n}{t} \right)\,\lambda_{n,\alpha}\,. \eeqa
Here, we have used the recursion relation, eq. (\ref{vk_recursive}), to get the second equality and the orthogonality condition to get the next step. The latter ensures that the last term in the second line vanishes since $N \geq 2(n+1)-1 = 2n+1$. The last equality results from eq. (\ref{nn_rel}). Following the identical steps, we can also show that
\beqa \label{col_rel}
\lambda_{\alpha,n+1} &=& \sum_k\,\frac{1}{m_k}\,v_{\alpha}^{(k)}\, v_{N+1-(n+1)}^{(k)}\, \nonumber \\ 
&=& \sum_k\,\frac{1}{m_k}\,v_{\alpha}^{(k)}\, \left[-v_{N+1-(n-1)}^{(k)}-\left(\frac{\epsilon_{N+1-n}-m_k}{t} \right) v_{N+1-n}^{(k)}\right]\, \nonumber \\  
&=& -\lambda_{\alpha,n-1} - \frac{\epsilon_{N+1-n}}{t}\,\lambda_{\alpha,n}\, \nonumber \\  
&=&  -\left(\kappa^\prime + \frac{\epsilon_{N+1-n}}{t} \right)\,\lambda_{\alpha,n}\,. \eeqa
Note that the above holds for $\alpha = 1,...,n$.

Finally, consider the remaining element in the last row which is $\lambda_{n+1,n+1}$. It can be related to $\lambda_{n,n+1}$ in the following way.
\beqa \label{diag_rel}
\lambda_{n+1,n+1} &=& \sum_k\,\frac{1}{m_k}\,v_{n+1}^{(k)}\, v_{N+1-(n+1)}^{(k)}\, \nonumber \\ 
&=& \sum_k\,\frac{1}{m_k}\, \left[-v_{n-1}^{(k)}-\left(\frac{\epsilon_{n}-m_k}{t} \right) v_{n}^{(k)}\right]\,v_{N+1-(n+1)}^{(k)}\, \nonumber \\  
&=& -\lambda_{n-1,n+1} - \frac{\epsilon_{n}}{t}\,\lambda_{n,n+1}\, \nonumber \\  
&=&  \left(\kappa^\prime + \frac{\epsilon_{N+1-n}}{t} \right)\,\lambda_{n-1,n}- \frac{\epsilon_{n}}{t}\,\lambda_{n,n+1}\, \nonumber \\  
&=&  \left(\kappa^\prime + \frac{\epsilon_{N+1-n}}{t} \right)\,\kappa\,\lambda_{n,n}- \frac{\epsilon_{n}}{t}\,\lambda_{n,n+1}\, \nonumber \\  
&=&  -\left(\kappa + \frac{\epsilon_n}{t} \right)\, \lambda_{n,n+1}\,. \eeqa
The steps in the first two lines are as usual. The smallest value $N+1-(n+1)$ can take, under the condition $N \geq (2 N_f-1)$, is $n+1$, and therefore the last term in the second line vanishes due to orthogonality of the eigenvectors. To get the expression in the fourth and the last line, we use the result, eq. (\ref{col_rel}).  In the second-last line, we have made use of eq. (\ref{nn_rel}). The final result along with eq. (\ref{row_rel}) implies that the $(n+1)^{\rm th}$ row is also a scaled version of the $n^{\rm th}$ row. Hence, $\lambda$ has only one independent row. This fact when substituted in eq. (\ref{Meff_def_app}) proves that ${\rm Rank} (m_{\rm eff})=1$. 

\section{Composition of the massless modes}
\label{app:massless_comp}
In this Appendix, we compute the eigenvectors corresponding to the massless modes of the tree-level mass matrix ${\cal M}$ when $N \geq (2N_f-1)$. The eigenvalue equation for the $(N_f-1)$ massless modes can be characterised as
\be \label{appB_ev_eq}
{\cal M}\,\Omega_R = 0\,,
\ee
with
\be  \label{appB_omega}
\Omega_R = \begin{pmatrix}
u_R \\ w_R \end{pmatrix}\,, 
\ee
where $u_R$ and $w_R$ are $N_f$- and $N$-dimensional column vectors, respectively. Using the form of ${\cal M}$ given in eq.~(\ref{M_fh}), one finds
\beqa \label{appB_ev_eq_2}
\mu_L\,w_R = 0\,,~~~~\mu_R\,u_R + M\,w_R =0\,.
\eeqa 
The first of the above equations, along with the fact that the first $N_f \times N_f$ block of $\mu_L$ is non-vanishing in general (see eq.~(\ref{muLR_fh})), leads to
\be \label{appB_v0_1}
w_{R,1} = w_{R,2} = ... =w_{R,N_f}=0\,.
\ee
The remaining $N-N_f$ elements of $w_R$ are non-vanishing in general at this stage.

The second equation in eq.~(\ref{appB_ev_eq_2}) implies
\be \label{appB_second_cond}
\sum_{\alpha=1}^{N_f} \left(\mu_R \right)_{i \alpha}\,u_{R,\alpha} + \epsilon_i\, w_{R,i} +t\, w_{R,{i+1}} + t\, w_{R,{i-1}} = 0\,,
\ee
for each $i=1,...,N$, where we have used eq.~(\ref{M_FC}). Since $\left(\mu_R \right)_{i \alpha}=0$ for $i = 1,...,N-N_f$ (see eq.~(\ref{muLR_fh})), one finds 
\be \label{appB_sec_cond}
\epsilon_i\, w_{R,i} +t\, w_{R,{i+1}} + t\, w_{R,{i-1}} = 0\,,
\ee
for each $i=1,...,N-N_f$. Note that the first $N_f$ elements of $w_R$ are already vanishing. Therefore, the above equation leads to
\be \label{appB_v0_2}
w_{R,{N_f+1}} = w_{R,{N_f+2}} = ... = w_{R,{N+1-N_f}}=0\,.
\ee

Using the results in eqs.~(\ref{appB_v0_1},\ref{appB_v0_2}), the $(N+N_f)$-dimensional eigenvector $\Omega_R$ can be parametrised as
\be \label{appB_OR}
\Omega_R = \left(u_{R,1},u_{R,2},...,u_{R,{N_f}},0,0,...,0,w_{R,{N+2-N_f}},w_{R,{N+3-N_f}},...,w_{R,N} \right)\,.
\ee
A similar analysis of the eigenvalue equation ${\cal M}^\dagger \Omega_L =0$ suggests that $\Omega_L$ has the form
\be \label{appB_OL}
\Omega_L = \left(u_{L,1},u_{L,2},...,u_{L,{N_f}},w_{L,1},w_{L,2},...,w_{L,{N_f-1}},0,0,...,0 \right)\,.
\ee
The magnitudes of the non-vanishing elements of $\Omega_{R(L)}$ depend on the explicit values of $\mu_{R(L)}$, $\epsilon_i$, and $t$. Numerically, we find that these elements take ${\cal O}(1)$ values if no hierarchy is assumed in the non-vanishing elements of $\mu_{R,L}$. The expressions in eqs.~(\ref{appB_OR},\ref{appB_OL}) imply that the $N_f-1$ massless modes are, in general, linear combinations of only $2N_f-1$ fields.

\bibliography{references}

\begin{thebibliography}{39}%
\makeatletter
\providecommand \@ifxundefined [1]{%
 \@ifx{#1\undefined}
}%
\providecommand \@ifnum [1]{%
 \ifnum #1\expandafter \@firstoftwo
 \else \expandafter \@secondoftwo
 \fi
}%
\providecommand \@ifx [1]{%
 \ifx #1\expandafter \@firstoftwo
 \else \expandafter \@secondoftwo
 \fi
}%
\providecommand \natexlab [1]{#1}%
\providecommand \enquote  [1]{``#1''}%
\providecommand \bibnamefont  [1]{#1}%
\providecommand \bibfnamefont [1]{#1}%
\providecommand \citenamefont [1]{#1}%
\providecommand \href@noop [0]{\@secondoftwo}%
\providecommand \href [0]{\begingroup \@sanitize@url \@href}%
\providecommand \@href[1]{\@@startlink{#1}\@@href}%
\providecommand \@@href[1]{\endgroup#1\@@endlink}%
\providecommand \@sanitize@url [0]{\catcode `\\12\catcode `\$12\catcode
  `\&12\catcode `\#12\catcode `\^12\catcode `\_12\catcode `\%12\relax}%
\providecommand \@@startlink[1]{}%
\providecommand \@@endlink[0]{}%
\providecommand \url  [0]{\begingroup\@sanitize@url \@url }%
\providecommand \@url [1]{\endgroup\@href {#1}{\urlprefix }}%
\providecommand \urlprefix  [0]{URL }%
\providecommand \Eprint [0]{\href }%
\providecommand \doibase [0]{http://dx.doi.org/}%
\providecommand \selectlanguage [0]{\@gobble}%
\providecommand \bibinfo  [0]{\@secondoftwo}%
\providecommand \bibfield  [0]{\@secondoftwo}%
\providecommand \translation [1]{[#1]}%
\providecommand \BibitemOpen [0]{}%
\providecommand \bibitemStop [0]{}%
\providecommand \bibitemNoStop [0]{.\EOS\space}%
\providecommand \EOS [0]{\spacefactor3000\relax}%
\providecommand \BibitemShut  [1]{\csname bibitem#1\endcsname}%
\let\auto@bib@innerbib\@empty
\bibitem [{\citenamefont {Dirac}(1937)}]{Dirac:1937ti}%
  \BibitemOpen
  \bibfield  {author} {\bibinfo {author} {\bibfnamefont {Paul A.~M.}\
  \bibnamefont {Dirac}},\ }\bibfield  {title} {\enquote {\bibinfo {title} {{The
  Cosmological constants}},}\ }\href {\doibase 10.1038/139323a0} {\bibfield
  {journal} {\bibinfo  {journal} {Nature}\ }\textbf {\bibinfo {volume} {139}},\
  \bibinfo {pages} {323} (\bibinfo {year} {1937})}\BibitemShut {NoStop}%
\bibitem [{\citenamefont {Weinberg}(1976)}]{Weinberg:1975gm}%
  \BibitemOpen
  \bibfield  {author} {\bibinfo {author} {\bibfnamefont {Steven}\ \bibnamefont
  {Weinberg}},\ }\bibfield  {title} {\enquote {\bibinfo {title} {{Implications
  of Dynamical Symmetry Breaking}},}\ }\href {\doibase
  10.1103/PhysRevD.19.1277} {\bibfield  {journal} {\bibinfo  {journal} {Phys.
  Rev. D}\ }\textbf {\bibinfo {volume} {13}},\ \bibinfo {pages} {974--996}
  (\bibinfo {year} {1976})},\ \bibinfo {note} {[Addendum: Phys.Rev.D 19,
  1277--1280 (1979)]}\BibitemShut {NoStop}%
\bibitem [{\citenamefont {Susskind}(1979)}]{Susskind:1978ms}%
  \BibitemOpen
  \bibfield  {author} {\bibinfo {author} {\bibfnamefont {Leonard}\ \bibnamefont
  {Susskind}},\ }\bibfield  {title} {\enquote {\bibinfo {title} {{Dynamics of
  Spontaneous Symmetry Breaking in the Weinberg-Salam Theory}},}\ }\href
  {\doibase 10.1103/PhysRevD.20.2619} {\bibfield  {journal} {\bibinfo
  {journal} {Phys. Rev. D}\ }\textbf {\bibinfo {volume} {20}},\ \bibinfo
  {pages} {2619--2625} (\bibinfo {year} {1979})}\BibitemShut {NoStop}%
\bibitem [{\citenamefont {Grossman}\ and\ \citenamefont
  {Neubert}(2000)}]{Grossman:1999ra}%
  \BibitemOpen
  \bibfield  {author} {\bibinfo {author} {\bibfnamefont {Yuval}\ \bibnamefont
  {Grossman}}\ and\ \bibinfo {author} {\bibfnamefont {Matthias}\ \bibnamefont
  {Neubert}},\ }\bibfield  {title} {\enquote {\bibinfo {title} {{Neutrino
  masses and mixings in nonfactorizable geometry}},}\ }\href {\doibase
  10.1016/S0370-2693(00)00054-X} {\bibfield  {journal} {\bibinfo  {journal}
  {Phys. Lett. B}\ }\textbf {\bibinfo {volume} {474}},\ \bibinfo {pages}
  {361--371} (\bibinfo {year} {2000})},\ \Eprint
  {http://arxiv.org/abs/hep-ph/9912408} {arXiv:hep-ph/9912408} \BibitemShut
  {NoStop}%
\bibitem [{\citenamefont {Gherghetta}\ and\ \citenamefont
  {Pomarol}(2000)}]{Gherghetta:2000qt}%
  \BibitemOpen
  \bibfield  {author} {\bibinfo {author} {\bibfnamefont {Tony}\ \bibnamefont
  {Gherghetta}}\ and\ \bibinfo {author} {\bibfnamefont {Alex}\ \bibnamefont
  {Pomarol}},\ }\bibfield  {title} {\enquote {\bibinfo {title} {{Bulk fields
  and supersymmetry in a slice of AdS}},}\ }\href {\doibase
  10.1016/S0550-3213(00)00392-8} {\bibfield  {journal} {\bibinfo  {journal}
  {Nucl. Phys. B}\ }\textbf {\bibinfo {volume} {586}},\ \bibinfo {pages}
  {141--162} (\bibinfo {year} {2000})},\ \Eprint
  {http://arxiv.org/abs/hep-ph/0003129} {arXiv:hep-ph/0003129} \BibitemShut
  {NoStop}%
\bibitem [{\citenamefont {Abe}\ \emph {et~al.}(2003)\citenamefont {Abe},
  \citenamefont {Kobayashi}, \citenamefont {Maru},\ and\ \citenamefont
  {Yoshioka}}]{Abe:2002rj}%
  \BibitemOpen
  \bibfield  {author} {\bibinfo {author} {\bibfnamefont {Hiroyuki}\
  \bibnamefont {Abe}}, \bibinfo {author} {\bibfnamefont {Tatsuo}\ \bibnamefont
  {Kobayashi}}, \bibinfo {author} {\bibfnamefont {Nobuhito}\ \bibnamefont
  {Maru}}, \ and\ \bibinfo {author} {\bibfnamefont {Koichi}\ \bibnamefont
  {Yoshioka}},\ }\bibfield  {title} {\enquote {\bibinfo {title} {{Field
  localization in warped gauge theories}},}\ }\href {\doibase
  10.1103/PhysRevD.67.045019} {\bibfield  {journal} {\bibinfo  {journal} {Phys.
  Rev. D}\ }\textbf {\bibinfo {volume} {67}},\ \bibinfo {pages} {045019}
  (\bibinfo {year} {2003})},\ \Eprint {http://arxiv.org/abs/hep-ph/0205344}
  {arXiv:hep-ph/0205344} \BibitemShut {NoStop}%
\bibitem [{\citenamefont {Falkowski}\ and\ \citenamefont
  {Kim}(2002)}]{Falkowski:2002cm}%
  \BibitemOpen
  \bibfield  {author} {\bibinfo {author} {\bibfnamefont {Adam}\ \bibnamefont
  {Falkowski}}\ and\ \bibinfo {author} {\bibfnamefont {Hyung~Do}\ \bibnamefont
  {Kim}},\ }\bibfield  {title} {\enquote {\bibinfo {title} {{Running of gauge
  couplings in AdS(5) via deconstruction}},}\ }\href {\doibase
  10.1088/1126-6708/2002/08/052} {\bibfield  {journal} {\bibinfo  {journal}
  {JHEP}\ }\textbf {\bibinfo {volume} {08}},\ \bibinfo {pages} {052} (\bibinfo
  {year} {2002})},\ \Eprint {http://arxiv.org/abs/hep-ph/0208058}
  {arXiv:hep-ph/0208058} \BibitemShut {NoStop}%
\bibitem [{\citenamefont {Randall}\ \emph {et~al.}(2003)\citenamefont
  {Randall}, \citenamefont {Shadmi},\ and\ \citenamefont
  {Weiner}}]{Randall:2002qr}%
  \BibitemOpen
  \bibfield  {author} {\bibinfo {author} {\bibfnamefont {Lisa}\ \bibnamefont
  {Randall}}, \bibinfo {author} {\bibfnamefont {Yael}\ \bibnamefont {Shadmi}},
  \ and\ \bibinfo {author} {\bibfnamefont {Neal}\ \bibnamefont {Weiner}},\
  }\bibfield  {title} {\enquote {\bibinfo {title} {{Deconstruction and gauge
  theories in AdS(5)}},}\ }\href {\doibase 10.1088/1126-6708/2003/01/055}
  {\bibfield  {journal} {\bibinfo  {journal} {JHEP}\ }\textbf {\bibinfo
  {volume} {01}},\ \bibinfo {pages} {055} (\bibinfo {year} {2003})},\ \Eprint
  {http://arxiv.org/abs/hep-th/0208120} {arXiv:hep-th/0208120} \BibitemShut
  {NoStop}%
\bibitem [{\citenamefont {Arkani-Hamed}\ \emph
  {et~al.}(2001{\natexlab{a}})\citenamefont {Arkani-Hamed}, \citenamefont
  {Cohen},\ and\ \citenamefont {Georgi}}]{Arkani-Hamed:2001nha}%
  \BibitemOpen
  \bibfield  {author} {\bibinfo {author} {\bibfnamefont {Nima}\ \bibnamefont
  {Arkani-Hamed}}, \bibinfo {author} {\bibfnamefont {Andrew~G.}\ \bibnamefont
  {Cohen}}, \ and\ \bibinfo {author} {\bibfnamefont {Howard}\ \bibnamefont
  {Georgi}},\ }\bibfield  {title} {\enquote {\bibinfo {title} {{Electroweak
  symmetry breaking from dimensional deconstruction}},}\ }\href {\doibase
  10.1016/S0370-2693(01)00741-9} {\bibfield  {journal} {\bibinfo  {journal}
  {Phys. Lett. B}\ }\textbf {\bibinfo {volume} {513}},\ \bibinfo {pages}
  {232--240} (\bibinfo {year} {2001}{\natexlab{a}})},\ \Eprint
  {http://arxiv.org/abs/hep-ph/0105239} {arXiv:hep-ph/0105239} \BibitemShut
  {NoStop}%
\bibitem [{\citenamefont {Arkani-Hamed}\ \emph
  {et~al.}(2001{\natexlab{b}})\citenamefont {Arkani-Hamed}, \citenamefont
  {Cohen},\ and\ \citenamefont {Georgi}}]{Arkani-Hamed:2001kyx}%
  \BibitemOpen
  \bibfield  {author} {\bibinfo {author} {\bibfnamefont {Nima}\ \bibnamefont
  {Arkani-Hamed}}, \bibinfo {author} {\bibfnamefont {Andrew~G.}\ \bibnamefont
  {Cohen}}, \ and\ \bibinfo {author} {\bibfnamefont {Howard}\ \bibnamefont
  {Georgi}},\ }\bibfield  {title} {\enquote {\bibinfo {title}
  {{(De)constructing dimensions}},}\ }\href {\doibase
  10.1103/PhysRevLett.86.4757} {\bibfield  {journal} {\bibinfo  {journal}
  {Phys. Rev. Lett.}\ }\textbf {\bibinfo {volume} {86}},\ \bibinfo {pages}
  {4757--4761} (\bibinfo {year} {2001}{\natexlab{b}})},\ \Eprint
  {http://arxiv.org/abs/hep-th/0104005} {arXiv:hep-th/0104005} \BibitemShut
  {NoStop}%
\bibitem [{\citenamefont {Hill}\ \emph {et~al.}(2001)\citenamefont {Hill},
  \citenamefont {Pokorski},\ and\ \citenamefont {Wang}}]{Hill:2000mu}%
  \BibitemOpen
  \bibfield  {author} {\bibinfo {author} {\bibfnamefont {Christopher~T.}\
  \bibnamefont {Hill}}, \bibinfo {author} {\bibfnamefont {Stefan}\ \bibnamefont
  {Pokorski}}, \ and\ \bibinfo {author} {\bibfnamefont {Jing}\ \bibnamefont
  {Wang}},\ }\bibfield  {title} {\enquote {\bibinfo {title} {{Gauge Invariant
  Effective Lagrangian for Kaluza-Klein Modes}},}\ }\href {\doibase
  10.1103/PhysRevD.64.105005} {\bibfield  {journal} {\bibinfo  {journal} {Phys.
  Rev. D}\ }\textbf {\bibinfo {volume} {64}},\ \bibinfo {pages} {105005}
  (\bibinfo {year} {2001})},\ \Eprint {http://arxiv.org/abs/hep-th/0104035}
  {arXiv:hep-th/0104035} \BibitemShut {NoStop}%
\bibitem [{\citenamefont {Choi}\ \emph {et~al.}(2014)\citenamefont {Choi},
  \citenamefont {Kim},\ and\ \citenamefont {Yun}}]{Choi:2014rja}%
  \BibitemOpen
  \bibfield  {author} {\bibinfo {author} {\bibfnamefont {Kiwoon}\ \bibnamefont
  {Choi}}, \bibinfo {author} {\bibfnamefont {Hyungjin}\ \bibnamefont {Kim}}, \
  and\ \bibinfo {author} {\bibfnamefont {Seokhoon}\ \bibnamefont {Yun}},\
  }\bibfield  {title} {\enquote {\bibinfo {title} {{Natural inflation with
  multiple sub-Planckian axions}},}\ }\href {\doibase
  10.1103/PhysRevD.90.023545} {\bibfield  {journal} {\bibinfo  {journal} {Phys.
  Rev. D}\ }\textbf {\bibinfo {volume} {90}},\ \bibinfo {pages} {023545}
  (\bibinfo {year} {2014})},\ \Eprint {http://arxiv.org/abs/1404.6209}
  {arXiv:1404.6209 [hep-th]} \BibitemShut {NoStop}%
\bibitem [{\citenamefont {Giudice}\ and\ \citenamefont
  {McCullough}(2017)}]{Giudice:2016yja}%
  \BibitemOpen
  \bibfield  {author} {\bibinfo {author} {\bibfnamefont {Gian~F.}\ \bibnamefont
  {Giudice}}\ and\ \bibinfo {author} {\bibfnamefont {Matthew}\ \bibnamefont
  {McCullough}},\ }\bibfield  {title} {\enquote {\bibinfo {title} {{A Clockwork
  Theory}},}\ }\href {\doibase 10.1007/JHEP02(2017)036} {\bibfield  {journal}
  {\bibinfo  {journal} {JHEP}\ }\textbf {\bibinfo {volume} {02}},\ \bibinfo
  {pages} {036} (\bibinfo {year} {2017})},\ \Eprint
  {http://arxiv.org/abs/1610.07962} {arXiv:1610.07962 [hep-ph]} \BibitemShut
  {NoStop}%
\bibitem [{\citenamefont {Kaplan}\ and\ \citenamefont
  {Rattazzi}(2016)}]{Kaplan:2015fuy}%
  \BibitemOpen
  \bibfield  {author} {\bibinfo {author} {\bibfnamefont {David~E.}\
  \bibnamefont {Kaplan}}\ and\ \bibinfo {author} {\bibfnamefont {Riccardo}\
  \bibnamefont {Rattazzi}},\ }\bibfield  {title} {\enquote {\bibinfo {title}
  {{Large field excursions and approximate discrete symmetries from a clockwork
  axion}},}\ }\href {\doibase 10.1103/PhysRevD.93.085007} {\bibfield  {journal}
  {\bibinfo  {journal} {Phys. Rev. D}\ }\textbf {\bibinfo {volume} {93}},\
  \bibinfo {pages} {085007} (\bibinfo {year} {2016})},\ \Eprint
  {http://arxiv.org/abs/1511.01827} {arXiv:1511.01827 [hep-ph]} \BibitemShut
  {NoStop}%
\bibitem [{\citenamefont {Craig}\ and\ \citenamefont
  {Sutherland}(2018)}]{Craig:2017ppp}%
  \BibitemOpen
  \bibfield  {author} {\bibinfo {author} {\bibfnamefont {Nathaniel}\
  \bibnamefont {Craig}}\ and\ \bibinfo {author} {\bibfnamefont {Dave}\
  \bibnamefont {Sutherland}},\ }\bibfield  {title} {\enquote {\bibinfo {title}
  {{Exponential Hierarchies from Anderson Localization in Theory Space}},}\
  }\href {\doibase 10.1103/PhysRevLett.120.221802} {\bibfield  {journal}
  {\bibinfo  {journal} {Phys. Rev. Lett.}\ }\textbf {\bibinfo {volume} {120}},\
  \bibinfo {pages} {221802} (\bibinfo {year} {2018})},\ \Eprint
  {http://arxiv.org/abs/1710.01354} {arXiv:1710.01354 [hep-ph]} \BibitemShut
  {NoStop}%
\bibitem [{\citenamefont {Rothstein}(2013)}]{Rothstein:2012hk}%
  \BibitemOpen
  \bibfield  {author} {\bibinfo {author} {\bibfnamefont {Ira~Z.}\ \bibnamefont
  {Rothstein}},\ }\bibfield  {title} {\enquote {\bibinfo {title}
  {{Gravitational Anderson Localization}},}\ }\href {\doibase
  10.1103/PhysRevLett.110.011601} {\bibfield  {journal} {\bibinfo  {journal}
  {Phys. Rev. Lett.}\ }\textbf {\bibinfo {volume} {110}},\ \bibinfo {pages}
  {011601} (\bibinfo {year} {2013})},\ \Eprint {http://arxiv.org/abs/1211.7149}
  {arXiv:1211.7149 [hep-th]} \BibitemShut {NoStop}%
\bibitem [{\citenamefont {Anderson}(1958)}]{Anderson:1958vr}%
  \BibitemOpen
  \bibfield  {author} {\bibinfo {author} {\bibfnamefont {P.~W.}\ \bibnamefont
  {Anderson}},\ }\bibfield  {title} {\enquote {\bibinfo {title} {{Absence of
  Diffusion in Certain Random Lattices}},}\ }\href {\doibase
  10.1103/PhysRev.109.1492} {\bibfield  {journal} {\bibinfo  {journal} {Phys.
  Rev.}\ }\textbf {\bibinfo {volume} {109}},\ \bibinfo {pages} {1492--1505}
  (\bibinfo {year} {1958})}\BibitemShut {NoStop}%
\bibitem [{\citenamefont {Georgi}(1986)}]{Georgi:1985hf}%
  \BibitemOpen
  \bibfield  {author} {\bibinfo {author} {\bibfnamefont {Howard}\ \bibnamefont
  {Georgi}},\ }\bibfield  {title} {\enquote {\bibinfo {title} {{A Tool Kit for
  Builders of Composite Models}},}\ }\href {\doibase
  10.1016/0550-3213(86)90092-1} {\bibfield  {journal} {\bibinfo  {journal}
  {Nucl. Phys. B}\ }\textbf {\bibinfo {volume} {266}},\ \bibinfo {pages}
  {274--284} (\bibinfo {year} {1986})}\BibitemShut {NoStop}%
\bibitem [{\citenamefont {Douglas}\ and\ \citenamefont
  {Moore}(1996)}]{Douglas:1996sw}%
  \BibitemOpen
  \bibfield  {author} {\bibinfo {author} {\bibfnamefont {Michael~R.}\
  \bibnamefont {Douglas}}\ and\ \bibinfo {author} {\bibfnamefont {Gregory~W.}\
  \bibnamefont {Moore}},\ }\bibfield  {title} {\enquote {\bibinfo {title}
  {{D-branes, quivers, and ALE instantons}},}\ }\href@noop {} {\  (\bibinfo
  {year} {1996})},\ \Eprint {http://arxiv.org/abs/hep-th/9603167}
  {arXiv:hep-th/9603167} \BibitemShut {NoStop}%
\bibitem [{\citenamefont {Izrailev}\ \emph {et~al.}(2012)\citenamefont
  {Izrailev}, \citenamefont {Krokhin},\ and\ \citenamefont
  {Makarov}}]{Izrailev_2012}%
  \BibitemOpen
  \bibfield  {author} {\bibinfo {author} {\bibfnamefont {F.M.}\ \bibnamefont
  {Izrailev}}, \bibinfo {author} {\bibfnamefont {A.A.}\ \bibnamefont
  {Krokhin}}, \ and\ \bibinfo {author} {\bibfnamefont {N.M.}\ \bibnamefont
  {Makarov}},\ }\bibfield  {title} {\enquote {\bibinfo {title} {Anomalous
  localization in low-dimensional systems with correlated disorder},}\ }\href
  {\doibase 10.1016/j.physrep.2011.11.002} {\bibfield  {journal} {\bibinfo
  {journal} {Physics Reports}\ }\textbf {\bibinfo {volume} {512}},\ \bibinfo
  {pages} {125–254} (\bibinfo {year} {2012})}\BibitemShut {NoStop}%
\bibitem [{\citenamefont {Economou}\ and\ \citenamefont
  {Cohen}(1971)}]{PhysRevB.4.396}%
  \BibitemOpen
  \bibfield  {author} {\bibinfo {author} {\bibfnamefont {E.~N.}\ \bibnamefont
  {Economou}}\ and\ \bibinfo {author} {\bibfnamefont {Morrel~H.}\ \bibnamefont
  {Cohen}},\ }\bibfield  {title} {\enquote {\bibinfo {title} {Localization in
  one-dimensional disordered systems},}\ }\href {\doibase
  10.1103/PhysRevB.4.396} {\bibfield  {journal} {\bibinfo  {journal} {Phys.
  Rev. B}\ }\textbf {\bibinfo {volume} {4}},\ \bibinfo {pages} {396--404}
  (\bibinfo {year} {1971})}\BibitemShut {NoStop}%
\bibitem [{\citenamefont {Pichard}(1986)}]{JLPichard_1986}%
  \BibitemOpen
  \bibfield  {author} {\bibinfo {author} {\bibfnamefont {J~L}\ \bibnamefont
  {Pichard}},\ }\bibfield  {title} {\enquote {\bibinfo {title} {The
  one-dimensional anderson model: scaling and resonances revisited},}\ }\href
  {\doibase 10.1088/0022-3719/19/10/009} {\bibfield  {journal} {\bibinfo
  {journal} {Journal of Physics C: Solid State Physics}\ }\textbf {\bibinfo
  {volume} {19}},\ \bibinfo {pages} {1519} (\bibinfo {year}
  {1986})}\BibitemShut {NoStop}%
\bibitem [{\citenamefont {Mohanta}\ and\ \citenamefont
  {Patel}(2022)}]{Mohanta:2022seo}%
  \BibitemOpen
  \bibfield  {author} {\bibinfo {author} {\bibfnamefont {Gurucharan}\
  \bibnamefont {Mohanta}}\ and\ \bibinfo {author} {\bibfnamefont {Ketan~M.}\
  \bibnamefont {Patel}},\ }\bibfield  {title} {\enquote {\bibinfo {title}
  {{Radiatively generated fermion mass hierarchy from flavor nonuniversal gauge
  symmetries}},}\ }\href {\doibase 10.1103/PhysRevD.106.075020} {\bibfield
  {journal} {\bibinfo  {journal} {Phys. Rev. D}\ }\textbf {\bibinfo {volume}
  {106}},\ \bibinfo {pages} {075020} (\bibinfo {year} {2022})},\ \Eprint
  {http://arxiv.org/abs/2207.10407} {arXiv:2207.10407 [hep-ph]} \BibitemShut
  {NoStop}%
\bibitem [{\citenamefont {Passarino}\ and\ \citenamefont
  {Veltman}(1979)}]{Passarino:1978jh}%
  \BibitemOpen
  \bibfield  {author} {\bibinfo {author} {\bibfnamefont {G.}~\bibnamefont
  {Passarino}}\ and\ \bibinfo {author} {\bibfnamefont {M.~J.~G.}\ \bibnamefont
  {Veltman}},\ }\bibfield  {title} {\enquote {\bibinfo {title} {{One Loop
  Corrections for e+ e- Annihilation Into mu+ mu- in the Weinberg Model}},}\
  }\href {\doibase 10.1016/0550-3213(79)90234-7} {\bibfield  {journal}
  {\bibinfo  {journal} {Nucl. Phys. B}\ }\textbf {\bibinfo {volume} {160}},\
  \bibinfo {pages} {151--207} (\bibinfo {year} {1979})}\BibitemShut {NoStop}%
\bibitem [{\citenamefont {Tropper}\ and\ \citenamefont
  {Fan}(2021)}]{Tropper:2020yew}%
  \BibitemOpen
  \bibfield  {author} {\bibinfo {author} {\bibfnamefont {Adam}\ \bibnamefont
  {Tropper}}\ and\ \bibinfo {author} {\bibfnamefont {Jiji}\ \bibnamefont
  {Fan}},\ }\bibfield  {title} {\enquote {\bibinfo {title}
  {{Randomness-Assisted Exponential Hierarchies}},}\ }\href {\doibase
  10.1103/PhysRevD.103.015001} {\bibfield  {journal} {\bibinfo  {journal}
  {Phys. Rev. D}\ }\textbf {\bibinfo {volume} {103}},\ \bibinfo {pages}
  {015001} (\bibinfo {year} {2021})},\ \Eprint
  {http://arxiv.org/abs/2001.07221} {arXiv:2001.07221 [hep-ph]} \BibitemShut
  {NoStop}%
\bibitem [{\citenamefont {Kaplan}(1991)}]{Kaplan:1991dc}%
  \BibitemOpen
  \bibfield  {author} {\bibinfo {author} {\bibfnamefont {David~B.}\
  \bibnamefont {Kaplan}},\ }\bibfield  {title} {\enquote {\bibinfo {title}
  {{Flavor at SSC energies: A New mechanism for dynamically generated fermion
  masses}},}\ }\href {\doibase 10.1016/S0550-3213(05)80021-5} {\bibfield
  {journal} {\bibinfo  {journal} {Nucl. Phys. B}\ }\textbf {\bibinfo {volume}
  {365}},\ \bibinfo {pages} {259--278} (\bibinfo {year} {1991})}\BibitemShut
  {NoStop}%
\bibitem [{\citenamefont {Babu}\ and\ \citenamefont
  {Mohapatra}(1993)}]{Babu:1992ia}%
  \BibitemOpen
  \bibfield  {author} {\bibinfo {author} {\bibfnamefont {K.~S.}\ \bibnamefont
  {Babu}}\ and\ \bibinfo {author} {\bibfnamefont {R.~N.}\ \bibnamefont
  {Mohapatra}},\ }\bibfield  {title} {\enquote {\bibinfo {title} {{Predictive
  neutrino spectrum in minimal SO(10) grand unification}},}\ }\href {\doibase
  10.1103/PhysRevLett.70.2845} {\bibfield  {journal} {\bibinfo  {journal}
  {Phys. Rev. Lett.}\ }\textbf {\bibinfo {volume} {70}},\ \bibinfo {pages}
  {2845--2848} (\bibinfo {year} {1993})},\ \Eprint
  {http://arxiv.org/abs/hep-ph/9209215} {arXiv:hep-ph/9209215} \BibitemShut
  {NoStop}%
\bibitem [{\citenamefont {Aulakh}\ \emph {et~al.}(2004)\citenamefont {Aulakh},
  \citenamefont {Bajc}, \citenamefont {Melfo}, \citenamefont {Senjanovic},\
  and\ \citenamefont {Vissani}}]{Aulakh:2003kg}%
  \BibitemOpen
  \bibfield  {author} {\bibinfo {author} {\bibfnamefont {Charanjit~S.}\
  \bibnamefont {Aulakh}}, \bibinfo {author} {\bibfnamefont {Borut}\
  \bibnamefont {Bajc}}, \bibinfo {author} {\bibfnamefont {Alejandra}\
  \bibnamefont {Melfo}}, \bibinfo {author} {\bibfnamefont {Goran}\ \bibnamefont
  {Senjanovic}}, \ and\ \bibinfo {author} {\bibfnamefont {Francesco}\
  \bibnamefont {Vissani}},\ }\bibfield  {title} {\enquote {\bibinfo {title}
  {{The Minimal supersymmetric grand unified theory}},}\ }\href {\doibase
  10.1016/j.physletb.2004.03.031} {\bibfield  {journal} {\bibinfo  {journal}
  {Phys. Lett. B}\ }\textbf {\bibinfo {volume} {588}},\ \bibinfo {pages}
  {196--202} (\bibinfo {year} {2004})},\ \Eprint
  {http://arxiv.org/abs/hep-ph/0306242} {arXiv:hep-ph/0306242} \BibitemShut
  {NoStop}%
\bibitem [{\citenamefont {Joshipura}\ and\ \citenamefont
  {Patel}(2011)}]{Joshipura:2011nn}%
  \BibitemOpen
  \bibfield  {author} {\bibinfo {author} {\bibfnamefont {Anjan~S.}\
  \bibnamefont {Joshipura}}\ and\ \bibinfo {author} {\bibfnamefont {Ketan~M.}\
  \bibnamefont {Patel}},\ }\bibfield  {title} {\enquote {\bibinfo {title}
  {{Fermion Masses in SO(10) Models}},}\ }\href {\doibase
  10.1103/PhysRevD.83.095002} {\bibfield  {journal} {\bibinfo  {journal} {Phys.
  Rev. D}\ }\textbf {\bibinfo {volume} {83}},\ \bibinfo {pages} {095002}
  (\bibinfo {year} {2011})},\ \Eprint {http://arxiv.org/abs/1102.5148}
  {arXiv:1102.5148 [hep-ph]} \BibitemShut {NoStop}%
\bibitem [{\citenamefont {Grisaru}\ \emph {et~al.}(1979)\citenamefont
  {Grisaru}, \citenamefont {Siegel},\ and\ \citenamefont
  {Rocek}}]{Grisaru:1979wc}%
  \BibitemOpen
  \bibfield  {author} {\bibinfo {author} {\bibfnamefont {Marcus~T.}\
  \bibnamefont {Grisaru}}, \bibinfo {author} {\bibfnamefont {W.}~\bibnamefont
  {Siegel}}, \ and\ \bibinfo {author} {\bibfnamefont {M.}~\bibnamefont
  {Rocek}},\ }\bibfield  {title} {\enquote {\bibinfo {title} {{Improved Methods
  for Supergraphs}},}\ }\href {\doibase 10.1016/0550-3213(79)90344-4}
  {\bibfield  {journal} {\bibinfo  {journal} {Nucl. Phys. B}\ }\textbf
  {\bibinfo {volume} {159}},\ \bibinfo {pages} {429} (\bibinfo {year}
  {1979})}\BibitemShut {NoStop}%
\bibitem [{\citenamefont {Seiberg}(1993)}]{Seiberg:1993vc}%
  \BibitemOpen
  \bibfield  {author} {\bibinfo {author} {\bibfnamefont {Nathan}\ \bibnamefont
  {Seiberg}},\ }\bibfield  {title} {\enquote {\bibinfo {title} {{Naturalness
  versus supersymmetric nonrenormalization theorems}},}\ }\href {\doibase
  10.1016/0370-2693(93)91541-T} {\bibfield  {journal} {\bibinfo  {journal}
  {Phys. Lett. B}\ }\textbf {\bibinfo {volume} {318}},\ \bibinfo {pages}
  {469--475} (\bibinfo {year} {1993})},\ \Eprint
  {http://arxiv.org/abs/hep-ph/9309335} {arXiv:hep-ph/9309335} \BibitemShut
  {NoStop}%
\bibitem [{\citenamefont {Dvali}\ and\ \citenamefont
  {Jankowsky}(2022)}]{Dvali:2020uvd}%
  \BibitemOpen
  \bibfield  {author} {\bibinfo {author} {\bibfnamefont {Gia}\ \bibnamefont
  {Dvali}}\ and\ \bibinfo {author} {\bibfnamefont {Anna}\ \bibnamefont
  {Jankowsky}},\ }\bibfield  {title} {\enquote {\bibinfo {title} {{Absence of
  the \ensuremath{\mu}-problem in grand unification}},}\ }\href {\doibase
  10.1103/PhysRevD.105.016009} {\bibfield  {journal} {\bibinfo  {journal}
  {Phys. Rev. D}\ }\textbf {\bibinfo {volume} {105}},\ \bibinfo {pages}
  {016009} (\bibinfo {year} {2022})},\ \Eprint
  {http://arxiv.org/abs/2009.07762} {arXiv:2009.07762 [hep-ph]} \BibitemShut
  {NoStop}%
\bibitem [{\citenamefont {Georgi}\ and\ \citenamefont
  {Glashow}(1973)}]{Georgi:1972hy}%
  \BibitemOpen
  \bibfield  {author} {\bibinfo {author} {\bibfnamefont {Howard}\ \bibnamefont
  {Georgi}}\ and\ \bibinfo {author} {\bibfnamefont {Sheldon~L.}\ \bibnamefont
  {Glashow}},\ }\bibfield  {title} {\enquote {\bibinfo {title} {{Attempts to
  calculate the electron mass}},}\ }\href {\doibase 10.1103/PhysRevD.7.2457}
  {\bibfield  {journal} {\bibinfo  {journal} {Phys. Rev. D}\ }\textbf {\bibinfo
  {volume} {7}},\ \bibinfo {pages} {2457--2463} (\bibinfo {year}
  {1973})}\BibitemShut {NoStop}%
\bibitem [{\citenamefont {Mohapatra}(1974)}]{Mohapatra:1974wk}%
  \BibitemOpen
  \bibfield  {author} {\bibinfo {author} {\bibfnamefont {Rabindra~N.}\
  \bibnamefont {Mohapatra}},\ }\bibfield  {title} {\enquote {\bibinfo {title}
  {{Gauge Model for Chiral Symmetry Breaking and Muon electron Mass Ratio}},}\
  }\href {\doibase 10.1103/PhysRevD.9.3461} {\bibfield  {journal} {\bibinfo
  {journal} {Phys. Rev. D}\ }\textbf {\bibinfo {volume} {9}},\ \bibinfo {pages}
  {3461} (\bibinfo {year} {1974})}\BibitemShut {NoStop}%
\bibitem [{\citenamefont {Barr}\ and\ \citenamefont {Zee}(1978)}]{Barr:1978rv}%
  \BibitemOpen
  \bibfield  {author} {\bibinfo {author} {\bibfnamefont {Stephen~M.}\
  \bibnamefont {Barr}}\ and\ \bibinfo {author} {\bibfnamefont {A.}~\bibnamefont
  {Zee}},\ }\bibfield  {title} {\enquote {\bibinfo {title} {{Calculating the
  Electron Mass in Terms of Measured Quantities}},}\ }\href {\doibase
  10.1103/PhysRevD.17.1854} {\bibfield  {journal} {\bibinfo  {journal} {Phys.
  Rev. D}\ }\textbf {\bibinfo {volume} {17}},\ \bibinfo {pages} {1854}
  (\bibinfo {year} {1978})}\BibitemShut {NoStop}%
\bibitem [{\citenamefont {Wilczek}\ and\ \citenamefont
  {Zee}(1979)}]{Wilczek:1978xi}%
  \BibitemOpen
  \bibfield  {author} {\bibinfo {author} {\bibfnamefont {Frank}\ \bibnamefont
  {Wilczek}}\ and\ \bibinfo {author} {\bibfnamefont {A.}~\bibnamefont {Zee}},\
  }\bibfield  {title} {\enquote {\bibinfo {title} {{Horizontal Interaction and
  Weak Mixing Angles}},}\ }\href {\doibase 10.1103/PhysRevLett.42.421}
  {\bibfield  {journal} {\bibinfo  {journal} {Phys. Rev. Lett.}\ }\textbf
  {\bibinfo {volume} {42}},\ \bibinfo {pages} {421} (\bibinfo {year}
  {1979})}\BibitemShut {NoStop}%
\bibitem [{\citenamefont {Yanagida}(1979)}]{Yanagida:1979gs}%
  \BibitemOpen
  \bibfield  {author} {\bibinfo {author} {\bibfnamefont {Tsutomu}\ \bibnamefont
  {Yanagida}},\ }\bibfield  {title} {\enquote {\bibinfo {title} {{Horizontal
  Symmetry and Mass of the Top Quark}},}\ }\href {\doibase
  10.1103/PhysRevD.20.2986} {\bibfield  {journal} {\bibinfo  {journal} {Phys.
  Rev. D}\ }\textbf {\bibinfo {volume} {20}},\ \bibinfo {pages} {2986}
  (\bibinfo {year} {1979})}\BibitemShut {NoStop}%
\bibitem [{\citenamefont {Barbieri}\ and\ \citenamefont
  {Nanopoulos}(1980)}]{Barbieri:1980tz}%
  \BibitemOpen
  \bibfield  {author} {\bibinfo {author} {\bibfnamefont {Riccardo}\
  \bibnamefont {Barbieri}}\ and\ \bibinfo {author} {\bibfnamefont {Dimitri~V.}\
  \bibnamefont {Nanopoulos}},\ }\bibfield  {title} {\enquote {\bibinfo {title}
  {{Hierarchical Fermion Masses From Grand Unification}},}\ }\href {\doibase
  10.1016/0370-2693(80)90395-0} {\bibfield  {journal} {\bibinfo  {journal}
  {Phys. Lett. B}\ }\textbf {\bibinfo {volume} {95}},\ \bibinfo {pages}
  {43--46} (\bibinfo {year} {1980})}\BibitemShut {NoStop}%
\bibitem [{\citenamefont {Balakrishna}(1988)}]{Balakrishna:1987qd}%
  \BibitemOpen
  \bibfield  {author} {\bibinfo {author} {\bibfnamefont {B.~S.}\ \bibnamefont
  {Balakrishna}},\ }\bibfield  {title} {\enquote {\bibinfo {title} {{Fermion
  Mass Hierarchy From Radiative Corrections}},}\ }\href {\doibase
  10.1103/PhysRevLett.60.1602} {\bibfield  {journal} {\bibinfo  {journal}
  {Phys. Rev. Lett.}\ }\textbf {\bibinfo {volume} {60}},\ \bibinfo {pages}
  {1602} (\bibinfo {year} {1988})}\BibitemShut {NoStop}%
\end{thebibliography}%
\end{document}